\documentclass[11pt]{article}

\usepackage{etoolbox}

\usepackage[utf8]{inputenc}

\usepackage[letterpaper,margin=1in]{geometry}

\usepackage{amsmath}
\usepackage{amssymb}

\usepackage{amsthm}

\usepackage{complexity}

\usepackage{dsfont}
\usepackage{bbm}

\usepackage[
   ruled,
   vlined,
   linesnumbered,
   boxed
]{algorithm2e}

\usepackage{hyperref}
\hypersetup{colorlinks, linkcolor=darkblue!60!blue, citecolor=darkgreen, urlcolor=darkblue}
\usepackage[capitalize, nameinlink]{cleveref}
\crefname{theorem}{Theorem}{Theorems}
\Crefname{lemma}{Lemma}{Lemmas}
\Crefname{claim}{Claim}{Claims}
\Crefname{fact}{Fact}{Facts}
\Crefname{remark}{Remark}{Remarks}
\Crefname{observation}{Observation}{Observations}
\Crefname{figure}{Figure}{Figures}
\Crefname{line}{Line}{Lines}
\Crefname{algocf}{Algorithm}{Algorithms}
\crefalias{AlgoLine}{line}
\Crefname{stepsromani}{Step}{Steps}
\Crefname{stepsarabici}{Step}{Steps}

\newtheorem{theorem}{Theorem}
\newtheorem{lemma}[theorem]{Lemma}

\newtheorem{proposition}[theorem]{Proposition}
\newtheorem{definition}[theorem]{Definition}

\newtheorem{corollary}[theorem]{Corollary}

\newtheorem{observation}[theorem]{Observation}

\usepackage{newtxtext, newtxmath}

\usepackage{anyfontsize}

\usepackage[inline]{enumitem}
\setlist[enumerate,1]{label=(\roman*), leftmargin=2.2em}
\setlist[enumerate,2]{label=(\alph*)}
\setlist{nosep,topsep=0.1em}
\setlist[itemize,1]{label={\bfseries--}}

\newlist{stepsroman}{enumerate}{1}
\setlist[stepsroman]{label=(\roman*), leftmargin=2.2em}

\newlist{stepsarabic}{enumerate}{1}
\setlist[stepsarabic]{rightmargin=0.2em, label=\arabic*., ref=\arabic*}

\usepackage{mathtools}

\usepackage{mdframed}
\mdfsetup{skipabove=\bigskipamount, skipbelow=\bigskipamount, innerleftmargin=0.5em, innertopmargin=0.5em, innerrightmargin=0.5em, innerbottommargin=0.5em, align=center}

\usepackage{xcolor}
\definecolor{darkblue}{rgb}{0,0,0.38}
\definecolor{darkred}{rgb}{0.6,0,0}
\definecolor{darkgreen}{rgb}{0.1,0.35,0}
\definecolor{or_green}{RGB}{100, 230,  0}
\definecolor{mid_red}{rgb}{1,0,0}
\definecolor{default_blue}{RGB}{0, 181, 226}
\definecolor{turkoise}{RGB}{125, 207, 182}
\definecolor{violet}{HTML}{982cc4}
\definecolor{mid_green}{rgb}{0, 0.8, 0.3}

\usepackage[color=darkred!60!white]{todonotes}
\setuptodonotes{size=\scriptsize}

\usepackage[
backend=biber,
style=alphabetic,
citestyle=alphabetic,
maxalphanames=6,
maxcitenames=99,
mincitenames=98,
maxbibnames=99,
giveninits=true,
url=false,
doi=true,
isbn=true,
backref=true
]{biblatex}

\newcommand{\footremember}[2]{\footnote{#2}
    \newcounter{#1}
    \setcounter{#1}{\value{footnote}}}

\makeatletter
 \newcommand{\linkdest}[1]{\Hy@raisedlink{\hypertarget{#1}{}}}
\makeatother
\newcommand{\BSTSO}{\protect\hyperlink{prb:BSTSO}{BSTSO}}
\newcommand{\vc}{\protect\hyperlink{prb:VertexCover}{vertex cover}}
\newcommand{\Vc}{\protect\hyperlink{prb:VertexCover}{Vertex cover}}

\makeatletter
\DeclareFieldFormat{eprint:arxiv}{arXiv\addcolon\space
  \ifhyperref
    {\href{https://arxiv.org/abs/#1}{\nolinkurl{#1}\iffieldundef{eprintclass}
     {}
{\addspace\mkbibbrackets{\thefield{eprintclass}}}}}
    {\nolinkurl{#1}
     \iffieldundef{eprintclass}
       {}
{\addspace\mkbibbrackets{\thefield{eprintclass}}}}}
\makeatother

\makeatletter
\newcommand{\labeltarget}[1]{\Hy@raisedlink{\hypertarget{#1}{}}}
\makeatother

\usepackage{wrapfig}

\usepackage[margin=10pt,font=small,labelfont=bf]{caption}

\usepackage{subcaption}

\usepackage{tikz}

\SetCustomAlgoRuledWidth{\textwidth}

\usepackage{xfrac}
\usepackage{algpseudocode}
\usepackage{amsfonts}
\ExplSyntaxOn
\DeclareRestrictedTemplate { xfrac } { text } { math }
  {
    numerator-font      = \number \fam ,
    slash-symbol        = /            ,
    slash-symbol-font   = \number \fam ,
    denominator-font    = \number \fam ,
    scale-factor        = 0.7          ,
    scale-relative      = false        ,
    scaling             = true         ,
    denominator-bot-sep = 0 pt         ,
    math-mode           = true         ,
    phantom             = ( }
\DeclareInstance { xfrac } { mathdefault } { math }
  { numerator-top-sep = 0pt }
\ExplSyntaxOff

\newcommand{\Z}{\mathbb{Z}}
\newcommand{\alg}{\operatorname{alg}}
\newcommand{\OPT}{\operatorname{OPT}}
\newcommand{\opt}{\operatorname{opt}}
\newcommand{\T}{\mathcal{T}}
\newcommand{\eqwithref}[2]{\stackrel{\makebox[0mm]{\footnotesize #1}}{\makebox[10mm]{$#2$}}}
\newcommand{\eqnoref}[1]{\makebox[10mm]{$#1$}}
\newcommand{\discup}{\mathbin{\dot{\cup}}}
\newcommand{\AND}{\textsc{And}}
\newcommand{\OR}{\textsc{Or}}
\newcommand{\XOR}{\textsc{Xor}}
\newcommand{\symdiff}{\mathbin{\bigtriangleup}}

\newbool{short}
\setbool{short}{false}
\newbool{long}
\setbool{long}{true}

\usetikzlibrary{shapes.gates.logic.US,shapes.gates.logic.IEC}
\usetikzlibrary{fadings}
\usetikzlibrary{calc}
\usetikzlibrary{backgrounds}
\usetikzlibrary{intersections}
\usetikzlibrary{positioning}
\usetikzlibrary{shapes}
\usetikzlibrary{decorations.pathmorphing}
\usetikzlibrary{fit}
\usetikzlibrary{matrix}
\usetikzlibrary{decorations.pathreplacing}
\usetikzlibrary{decorations.markings}
\usetikzlibrary{shapes.misc, shapes.geometric}
\usetikzlibrary{shadows}
\usetikzlibrary{arrows.meta}
\usetikzlibrary{shadings}
\usetikzlibrary{matrix}
\usetikzlibrary{fadings}
\usetikzlibrary{calligraphy}
\usetikzlibrary{decorations.pathmorphing}
\usetikzlibrary{calc}
\tikzfading[name=fade out, inner color=transparent!0, outer color=transparent!100]

\newcommand{\rot}[1]{\rotatebox{90}{#1}}
\tikzset{and-gate/.style={fill=mid_red,outer sep=0pt, thick, and gate US, draw, rotate=270}}
\tikzset{or-gate/.style={fill=or_green,outer sep=0pt, thick, or gate US, draw, rotate=270}}
\tikzset{shade/.style={circle, draw=black, fill=white, minimum size=2mm}}

\tikzset{fadeout/.style={or_green, path fading=fade out}}

\title{Size Minimization For Multi-Output \AND-Functions}

\author{Susanne Armbruster\footremember{UBonnHCM}{Research Institute for Discrete Mathematics and Hausdorff Center for Mathematics, University of Bonn, Bonn, Germany.
      Email:
      \href{mailto:armbruster@or.uni-bonn.de}{armbruster@or.uni-bonn.de}}
}\date{}

\begin{document}

\maketitle

\begin{abstract}
Recent improvements in adder optimization could be achieved by optimizing the \AND-trees occurring within the constructed circuits.
The overlap of such trees and its potential for pure size optimization has not been taken into account though.
Motivated by this, we examine the fundamental problem of minimizing the size of a circuit for multiple \AND-functions on intersecting variable sets.
Our formulation generalizes the overlapping \AND-trees within adder optimization but is in NP, in contrast to general Boolean circuit optimization which is in~$\Sigma_2^p$ (and thus suspected not to be in NP).
While restructuring the \AND- or \XOR-trees simultaneously, we optimize the total number of gates needed for all functions to be computed.
We show that this problem is APX-hard already for functions of few variables and present efficient approximation algorithms for the case in which the Boolean functions depend on at most~3 or~4 variables each,
achieving guarantees of~$\sfrac 43$ and~$1.9$, respectively.
To conclude, we give a polynomial approximation algorithm with guarantee~$\sfrac 23k$ for \AND-functions of up to~$k$ variables.
To achieve these results, the key technique is to determine how much overlap among the variable sets makes tree construction cheap and how little makes the optimum solution large.
\end{abstract}

\bigskip

\thispagestyle{empty}
\addtocounter{page}{-1}

\newpage

\section{Introduction}\label{sec:introduction}
In this work, we examine the problem of constructing a small circuit for a symmetric Boolean function ${f \colon \{0,1\}^n \to \{0,1\}}^m$ (for~$n, m \in \Z_{>0}$)
in which each component is concatenated from an associative symmetric binary operator~${\circ \colon \{0,1\}^2 \to \{0,1\}}$,
i.e.,~$f_j(x_1, \dots, x_n) = \bigcirc_{i \in T_j \subseteq \{1, \dots , n\}} x_i$ for~$1 \leq j \leq m$.
The main examples are the \AND-function ${f(x_1, \dots, x_k) = x_1 \land \dots \land x_k}$, the \OR-function~$f(x_1, \dots, x_k) = x_1 \lor \dots \lor x_k$
and the \XOR-function $f(x_1, \dots, x_k) = \sum_{i=1}^k x_i \pmod 2$.
For such a function, we aim to find a formula computing~$f$ with the minimum number of operations.

The minimization of general Boolean circuits is~$\Sigma_2^p$-complete (see~\cite{buchfuhrer2011complexity}), making it an extremely challenging problem.
Consequently, researchers focus on optimizing easier subproblems.
Adders, crucial components on modern computer chips, have become a focal point for optimization efforts.
The recent improvement of~\cite{brenner_hermann_aop_restructuring} over~\cite{grinchuk2009sharpening} for developing the fastest adders known today could be achieved by enhancing \AND-trees within the constructed circuits.
Thus, we want to focus on the opportunities of optimizing multiple \AND-trees simultaneously.

Our problem formulation slightly generalizes the problem appearing within adders, offering a broader perspective while its decision version remains in the NP complexity class.
Like this, it not only includes the instances generated within adder optimization but also all \AND- respectively \XOR-trees that can be found all over a circuit after application of De Morgan rules.

Let us now formally define the problem.
A circuit for a function~$f$ as mentioned above is a graph-based model to compute the function from elementary building blocks called gates, each computing the symmetric operator on two variables.
Each vertex of the directed graph either corresponds to some gate and computes the binary operator on its two predecessors, or it is an input without predecessors, representing a variable.
For simplicity, our focus in this work is on circuits where the underlying undirected graph is acyclic.
This condition ensures that there is precisely one path from an input to a gate containing the variable associated with that input.
The results transfer to other circuits, too, however.
\begin{definition}
   A circuit~$\mathcal{C} = (\mathcal{V}, \mathcal{E})$ is a directed acyclic graph with vertex set~$\mathcal{V} = \mathcal{I} \discup \mathcal{G}$ and edge set~$\mathcal{E}$ satisfying the following conditions.
   \begin{enumerate}
      \item Every vertex~$i \in \mathcal{I}$ satisfies~$\delta^-(i) = \emptyset$, i.e., has no incoming edges, and is associated with some Boolean variable~$x_i$.
      These vertices are called inputs.
      \item Each vertex~$v \in \mathcal{G}$ satisfies~$\lvert \delta^-(v) \rvert = 2$, i.e.,~$v$ has exactly two incoming edges.
      These vertices are called gates.
      \item The underlying undirected graph is acyclic.
   \end{enumerate}
   The Boolean function~$g_v$ associated with~$v \in \mathcal{G}$ is recursively given by~$g_v = \bigcirc_{w \in \delta^-(v)} f_w$, where~$g_w = x_w$ for inputs~${w \in \mathcal{I}}$.
   A circuit computes a function~$f=(f_1, \dots, f_m)$ if for each component~$f_j$ there is a vertex~$v \in \mathcal{G}$ such that~$f_j = g_v$.
   These vertices are called outputs.
\end{definition}

Motivated from VLSI design, our objective function is to minimize the number of gates that we use to compute multiple such functions on possibly intersecting variable sets.
The number of gates, also called the size of the circuit, holds significance in chip design.
A reduced number of gates not only simplifies placing a physical copy of the circuit on a chip but typically also results in decreased routing congestion and reduced power consumption of such a chip, which are all critical metrics in chip design.

Apart from size, a common objective for logic optimization on chips is delay minimization.
It is often realized by minimizing the depth, i.e., the length of the longest path in the circuit graph.
We will not optimize this directly here.
However, our algorithms usually build circuits with depth at most 1 more than minimum and concentrate on restructuring instances with small depth anyway.
Our main problem formulation is as follows, using~$\mathcal{P}(X)$ for the power set of a base set~$X$.

\begin{mdframed}[userdefinedwidth=0.95\linewidth, nobreak=true]
{\textbf{Binary Symmetric Tree Size Optimization (\BSTSO{}\linkdest{prb:BSTSO}):}}
   Given a set of variables~$x_1, \dots, x_n$, a set of index subsets~$\T \subseteq \mathcal{P}(\{1, \dots, n\})$ and an associative symmetric operator~$\circ \colon \{0,1\}^2 \to \{0,1\}$,
   find a common circuit computing~$\bigcirc_{i \in T} x_i$ for each~$T \in \T$, minimizing the number of used gates.
\end{mdframed}

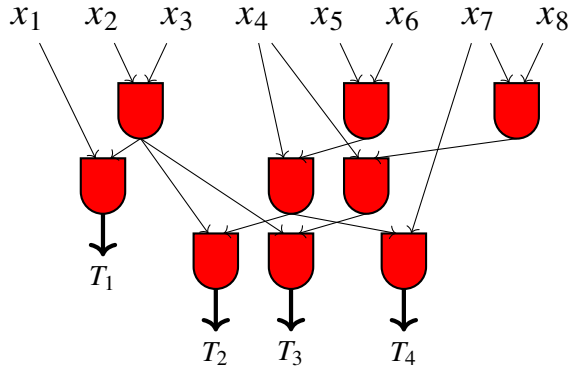
\begin{figure}[htbp]
   \begin{center}

      \begin{tikzpicture}[scale = 1.0]
         \node[anchor=center, align=center, text width=15cm] at (5,4.5) (penalties) {
            \begin{align*}
               \mathcal{T} = \left\{ T_1 = \{1, 2, 3\}, T_2 = \{2, 3, 4, 5, 6\}, T_3 = \{2, 3, 4, 7, 8\}, T_4 = \{4, 5, 6, 7 \}\right\}\end{align*}
         };

         \node[scale=1.3] (i1) at (1.5, 3.2){$x_1$};
         \node[scale=1.3] (i2) at (2.5, 3.2){$x_2$};
         \node[scale=1.3] (i3) at (3.5, 3.2){$x_3$};
         \node[scale=1.3] (i4) at (4.5, 3.2){$x_4$};
         \node[scale=1.3] (i5) at (5.5, 3.2){$x_5$};
         \node[scale=1.3] (i6) at (6.5, 3.2){$x_6$};
         \node[scale=1.3] (i7) at (7.5, 3.2){$x_7$};
         \node[scale=1.3] (i8) at (8.5, 3.2){$x_8$};

         \node[and-gate] (and1) at (3,2) {};
         \node[and-gate] (and2) at (2.5, 1) {};
         \node[] (output1) at (2.5, -0.25) {$T_1$};

         \node[and-gate] (and3) at (6,2) {};
         \node[and-gate] (and4) at (5,1) {};
         \node[and-gate] (and5) at (4,0) {};
         \node[] (output2) at (4, -1.25) {$T_2$};

         \node[and-gate] (and6) at (8,2) {};
         \node[and-gate] (and7) at (6,1) {};
         \node[and-gate] (and8) at (5,0) {};
         \node[] (output3) at (5, -1.25) {$T_3$};

         \node[and-gate] (and9) at (6.5, 0) {};
         \node[] (output4) at (6.5, -1.25) {$T_4$};

         \draw[->] (i2) -- (and1.input 2);
         \draw[->] (i3) -- (and1.input 1);
         \draw[->] (and1.output) -- (and2.input 1);
         \draw[->] (i1) -- (and2.input 2);
         \draw[ultra thick, ->] (and2.output) -- (output1);

         \draw[->] (i5) -- (and3.input 2);
         \draw[->] (i6) -- (and3.input 1);
         \draw[->] (and3.output) -- (and4.input 1);
         \draw[->] (i4) -- (and4.input 2);
         \draw[->] (and1.output) -- (and5.input 2);
         \draw[->] (and4.output) -- (and5.input 1);
         \draw[ultra thick, ->] (and5.output) -- (output2);

         \draw[->] (i7) -- (and6.input 2);
         \draw[->] (i8) -- (and6.input 1);
         \draw[->] (i4) -- (and7.input 2);
         \draw[->] (and6.output) -- (and7.input 1);
         \draw[->] (and7.output) -- (and8.input 1);
         \draw[->] (and1.output) -- (and8.input 2);
         \draw[ultra thick, ->] (and8.output) -- (output3);

         \draw[->] (i7) -- (and9.input 1);
         \draw[->] (and4.output) -- (and9.input 2);
         \draw[ultra thick, ->] (and9.output) -- (output4);

      \end{tikzpicture}
      \caption{Example instance~$\T$ with a possible implementation solving the \BSTSO\ problem.
      The inputs are shown as variables on the top.
      In \textcolor{mid_red}{red} you can see the \AND-gates.
      Thick arrows pointing into nowhere symbolize outputs of the circuit.
      They are associated with the indicated tree~$T$.}
      \label{fig:stso_example}
   \end{center}
\end{figure}
An instance of \BSTSO\ will usually be given by~$\T$, implying the set of variables as union of the elements of~$T$.
We refer to the elements of~$\T$, for which we aim to construct a subcircuit, as trees.
An example instance and a possible solution are depicted in \cref{fig:stso_example}.
The goal of building these trees is equivalent for all symmetric operators~$\circ$ and it is only important how to combine the different gates and inputs.
To minimize the size of the constructed circuit, the crucial question is how to decide which intermediate results are computed and which trees should share intermediate results for subsets of their intersecting variable sets.
As mentioned, we look for a circuit where the underlying undirected graph for a single~$T$ does not contain any cycles.
This holds true if the function is computed without redundancies, i.e., if each output is computed with the minimum number of operations.

\subsection{Related work}\label{subsec:former_work}
Various researchers have worked on optimizing single \AND- respectively \XOR-trees.
Such a single tree has a fixed size, and optimizing the delay (or depth) can easily be done by a dynamic program, as shown by~\cite{golumbic1976combinatorial} and
others like~\cite{parker_huffman} and~\cite{hoover_bounding_fanout} after him.
There are different approaches to optimize such trees in practice, including different objectives.
A common one is to consider the netlength of the nets incident to the instance in addition to timing.
\cite{kaemerling} show a two-level algorithm to optimize timing and include the netlength optimization of some side outputs, i.e., successors on a chip, of the inputs.
\cite{10.1145/1735023.1735046} give a method how to restructure trees for netlength-optimization only, using an already existing chip placement as orientation.

\AND-trees are often optimized as prefix graphs within adders.
\cite{kogge_stone_adder} presented a technique for computing so-called prefix graphs with minimum depth of~$\log_2 n$ and almost linear size.
\cite{krapchenko1970asymptotic} also showed how to construct a prefix graph, albeit with linear size and depth~$2 \log_2 n$ instead.
Regarding adders, this has been developed further by~\cite{spirkl_held}, who combine previous approaches to achieve a depth of~$\log_2 n$ under a fanout bound of 2, i.e., with at most two successors per vertex.

Since the first prefix graph approaches strategies have improved and the asymptotically fastest and smallest adders nowadays have been developed by~\cite{grinchuk2009sharpening} and improved by~\cite{brenner_hermann_aop_restructuring}.
To build these adders, the latest key improvement was to optimize \AND- respectively \OR-trees occurring as carry bits with respect to their depth and size.

This gives rise to the interesting question of how these trees can be optimized in general, without being restricted to the special instances arising in adder minimization.
We study a very general version of this question, namely \AND-trees on input sets with arbitrary overlap, which is however still in NP\@.
This contrasts minimizing general Boolean circuits, which is a~$\sum_2^p$-complete problem~\cite{buchfuhrer2011complexity}.
Regarding the optimization of several symmetric functions at once, only little has been done so far.
~\cite{christian_ma} shows that the problem of size optimization under minimum delay constraints is NP-hard.
We will see in \cref{sec:hardness} that this is even true without the delay constraints.

\subsection{Our results and contributions}\label{subsec:our_results}
In this paper we want to give some positive and negative results regarding approximation guarantees of the \BSTSO\ problem.
We call an algorithm an~$\alpha$-approximation if it runs in polynomial time and returns a solution having a size that is by a fraction of at most~$\alpha$ larger than the optimum size.
To quantify our results, we introduce~$k$ to be the maximum number of variables of a tree, i.e.,~$k \coloneqq \max\left\{\lvert T \rvert : T \in \T\right\}$.
We start by showing the following hardness result.

\begin{theorem}\label{thm:hardness_intro}
   There is some~$\alpha > 1$ such that there exists no~$\alpha$-approximation for \BSTSO\ unless P=NP.
   This is true even when restricting to instances with trees containing~$3$ variables each, i.e., $k = 3$.
\end{theorem}
We use a reduction of \vc\ to prove this.
This reduction is not approximation preserving but still allows us to give a lower bound on the guarantee of \BSTSO\ due to the hardness of \vc\ in graphs with bounded degree.
As this theorem already suggests that the hardness of \BSTSO\ increases with~$k$, we will prove approximation guarantees depending on~$k$.
Our main result are the following approximation guarantees.

\begin{theorem}\label{thm:main_result}
   Let~$k$ be the maximum size of a tree in a \BSTSO\ instance.
   \begin{enumerate}[label=(\roman*)]
      \item \label{itm:k_3} There exists an efficient~$\frac 43$-approximation for~$k = 3$.
      \item \label{itm:k_4} There exists an efficient~$1.9$-approximation for~$k = 4$.
      \item \label{itm:general_k} There exists an efficient~$\frac{2}{3}k$-approximation for general~$k$.
   \end{enumerate}
\end{theorem}

To derive our results, we mainly focus on finding sets of trees with large intersection.
If the trees in a set~$\T' \subseteq \T$ share a large number of their variables, we can build this intersection from scratch and in addition use the built intermediate result for each~$T \in \T'$.
Afterwards we consider the remaining instance separately.
For this instance, we can deduce that every fairly large set of trees can only have a small overlap among their variable sets.
That is why the remaining instance must have a large optimum solution compared to the number of remaining trees.
We will evaluate how large precisely the intersection should be and how many trees we should at least intersect.
While it turns out that this approach only provides a constant factor approximation for instances of small~$k$, it is still beneficial in the general case.

The algorithm for the general case of item~\ref{itm:general_k} generalizes the idea used to prove item~\ref{itm:k_3}, both solving parts of the instance with few gates and afterwards proving that the optimum solution for the remaining instance is rather large.
For item~\ref{itm:k_4}, the idea is essentially the same but the analysis gets more complicated as we distinguish different kinds of trees, depending on the number of outgoing edges of their predecessors.
This however pays back by achieving the best ratio of approximation guarantee and maximum tree size~$k$.

\subsection{Organization of the paper}\label{subsec:paper_overview}
In \cref{sec:hardness}, we give a proof of the NP-hardness of \BSTSO\ and an extension to show \cref{thm:hardness_intro}.
\cref{sec:three_variables} gives a~$\frac 43$-approximation for the case where each tree has at most three variables proving \cref{thm:main_result}~\ref{itm:k_3}.
In \cref{sec:four_variables}, we deal with the case of four variables, proving \cref{thm:main_result}~\ref{itm:k_4}.
\cref{sec:arbitrary_variables} then shows a proof for the general case, i.e. \cref{thm:main_result}~\ref{itm:general_k}.

\section{Hardness of the problem}\label{sec:hardness}
We start by giving a reduction of \vc\ to \BSTSO\@ to show its NP-hardness.
Since we later deal with the hypergraph variant of \vc, too, we repeat the problem definition here.

\begin{mdframed}[userdefinedwidth=0.95\linewidth]
   \textbf{\Vc{}\linkdest{prb:VertexCover}:}
   Given an undirected (hyper)graph~$G = (V, E)$, compute a minimum size subset~${C \subseteq V}$ of the vertices such that each edge~$e \in E$ is covered, i.e., $C \cap e \neq \emptyset$.
\end{mdframed}
A proof of the NP-hardness of \vc\ can be found in~\cite{Karp1972}.

\subsection{NP-hardness}\label{subsec:np_hardness}

\begin{theorem}\label{thm:np_hardness}
The \BSTSO\@ problem is NP-hard.
\end{theorem}
\begin{proof}
Given a \vc\ instance~$G = (V, E)$ where~$V = \{1, \dots, n\}$, we create the following \BSTSO\ instance.
Let there be~$n+1$ Boolean variables~$x_0, \dots, x_n$.
For each edge~$e = \{i,j\} \in E$, we create a tree~$T = \{0, i, j\}$ for all~$e \in E$.
\bigskip

\textbf{Claim:}
There is a vertex cover of size~$N$ if and only if there is a \BSTSO\ solution of size~$N+\lvert E \rvert$ in the above construction.

\begin{figure}[htbp]
   \begin{center}
      \begin{tikzpicture}[scale=0.8]
         \normalsize

         \newcommand{\xdist}{4}

         \begin{scope}[shift={(0,0)}, scale=1.0, every node/.append style={transform shape}]
            \fill[fadeout] (0,0) circle (0.7);
            \fill[fadeout] (2*\xdist,0.5*\xdist) circle (0.7);
            \fill[fadeout] (\xdist,\xdist) circle (0.7);

            \fill[fadeout] (\xdist + 3, -1.5) circle (0.7);
            \node[shade] at (\xdist + 3, -1.5) {\textcolor{white}{1}};
            \node[] at (2*\xdist +1, -1.5) (text) {\textcolor{or_green}{Vertex cover}};

            \node[shade] at (0,0) (1) {1};
            \node[shade] at (\xdist,0) (2) {2};
            \node[shade] at (2*\xdist,0.5*\xdist) (3) {3};
            \node[shade] at (\xdist,\xdist) (4) {4};
            \node[shade] at (0,\xdist) (5) {5};

            \draw (1) -- (2);
            \draw (2) -- (3);
            \draw (3) -- (4);
            \draw (4) -- (5);
            \draw (5) -- (1);
            \draw (2) -- (4);

            \node[label=right:{\color{default_blue}$\T = \left\{\{0, i, j\} : \{i,j\} \in E(G)\right\}$}] at (0, -1.5) (T) {};
            \draw (1) -- (2)node [midway, fill = white] {\color{default_blue}\small$\{0, 1, 2\}$};
            \draw (2) -- (3)node [midway, fill = white] {\color{default_blue}\small$\{0, 2, 3\}$};
            \draw (3) -- (4)node [midway, fill = white] {\color{default_blue}\small$\{0, 3, 4\}$};
            \draw (4) -- (5)node [midway, fill = white] {\color{default_blue}\small$\{0, 4, 5\}$};
            \draw (5) -- (1)node [midway, fill = white] {\color{default_blue}\small$\{0, 1, 5\}$};
            \draw (2) -- (4)node [midway, fill = white] {\color{default_blue}\small$\{0, 2, 4\}$};
         \end{scope}

         \draw[ultra thick, <->] (2 * \xdist + 2, 2) -- (2 * \xdist + 4, 2);

         \begin{scope}[shift={(\xdist * 2 + 4,1)}, scale=1.0, every node/.append style={transform shape}]
            \node[scale=1.3] (i0) at (1.5, 3.2){$x_0$};
            \node[scale=1.3] (i1) at (2.5, 3.2){$x_1$};
            \node[scale=1.3] (i2) at (3.5, 3.2){$x_2$};
            \node[scale=1.3] (i3) at (4.5, 3.2){$x_3$};
            \node[scale=1.3] (i4) at (5.5, 3.2){$x_4$};
            \node[scale=1.3] (i5) at (6.5, 3.2){$x_5$};

            \node[and-gate] (and1) at (2,2) {$\rot{1}$};
            \node[and-gate] (and4) at (3.5,2) {$\rot{3}$};
            \node[and-gate] (and7) at (4.5,2) {$\rot{4}$};

            \draw[->] (i0) -- (and1.input 2);
            \draw[->] (i1) -- (and1.input 1);
            \draw[->] (i0) -- (and4.input 2);
            \draw[->] (i3) -- (and4.input 1);
            \draw[->] (i0) -- (and7.input 2);
            \draw[->] (i4) -- (and7.input 1);

            \node[and-gate] (and2) at (2, 0.5) {};
            \node[] (output1) at (2, -2) {$\{0, 1, 2\}$};
            \node[and-gate] (and3) at (5.625,0.5) {};
            \node[] (output2) at (5.625, -2) {$\{0, 1, 5\}$};

            \node[and-gate] (and5) at (2.875,0.5) {};
            \node[] (output3) at (2.875,-1) {$\{0, 2, 3\}$};
            \node[and-gate] (and6) at (4.75,0.5) {};
            \node[] (output4) at (4.75, -1) {$\{0, 3, 4\}$};

            \node[and-gate] (and8) at (3.875,0.5) {};
            \node[] (output5) at (3.875, -2) {$\{0, 2, 4\}$};
            \node[and-gate] (and9) at (6.5,0.5) {};
            \node[] (output6) at (6.5, -1) {$\{0, 4, 5\}$};

            \draw[->] (and1.output) -- (and2.input 2);
            \draw[->] (i2) to[out=225, in=75] (and2.input 1);
            \draw[ultra thick, ->] (and2.output) -- (output1);

            \draw[->] (i5) -- (and3.input 1);
            \draw[->] (and1.output) -- (and3.input 2);
            \draw[ultra thick, ->] (and3.output) -- (output2);

            \draw[->] (i2) to[out=240, in=75]  (and5.input 2);
            \draw[->] (and4.output) -- (and5.input 1);
            \draw[ultra thick, ->] (and5.output) -- (output3);

            \draw[->] (i4) -- (and6.input 1);
            \draw[->] (and4.output) -- (and6.input 2);
            \draw[ultra thick, ->] (and6.output) -- (output4);

            \draw[->] (i2) to[out=300, in=85] (and8.input 2);
            \draw[->] (and7.output) -- (and8.input 1);
            \draw[ultra thick, ->] (and8.output) -- (output5);

            \draw[->] (i5) -- (and9.input 1);
            \draw[->] (and7.output) -- (and9.input 2);
            \draw[ultra thick, ->] (and9.output) -- (output6);
         \end{scope}
      \end{tikzpicture}
      \caption{Correspondence of a vertex cover of size~$3$ and a solution to \BSTSO\ of size~$8$.
      Given the graph shown in black on the left, we construct the trees corresponding to the graph edges as indicated in \textcolor{cyan}{blue}.
      If we are now given a \textcolor{or_green}{vertex cover} for the graph, we can construct a \textcolor{mid_red}{circuit} as on the right and vice versa.}
      \label{fig:vc_example}
   \end{center}
\end{figure}
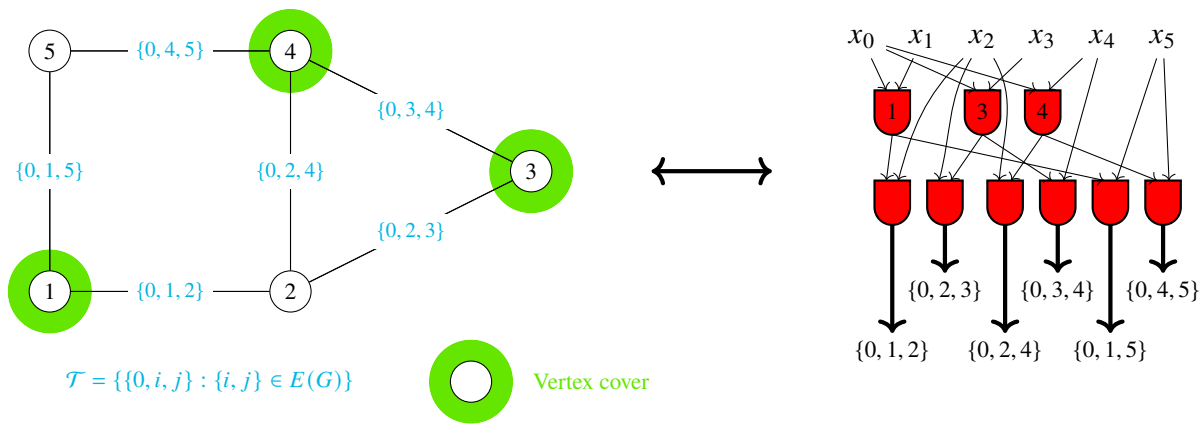
\bigskip

To prove this, assume you are given a vertex cover of size~$N$.
Then for each vertex~$i$ selected in the vertex cover, add a circuit node computing~$x_0 \circ x_i$, resulting in a total of~$N$ nodes.
Now each~$T \in \mathcal{T}$ is associated with a unique edge~$\{i,j\}$ for which it was introduced.
This edge is covered by one of its endpoints in the vertex cover, say without loss of generality~$i$.
Then there is already a circuit node~$x_0 \circ x_i$ and only one additional node is needed to compute~$x_j \circ (x_0 \circ x_i)$.
This makes a total of~$N + \lvert E \rvert$ nodes.

On the contrary, suppose you are given a solution to \BSTSO\ of value~$N + \lvert E \rvert$.
There are exactly~$\lvert E \rvert$ nodes in the circuit computing a Boolean function on three variables.
Thus, there remain~$N$ other nodes, which compute Boolean functions on two variables as intermediate results.
Since each edge~$e = \{i, j\}$ requires an intermediate result for the corresponding tree~$T = \{0, i, j\}$, there must be a node computing either~$x_0 \circ x_i$, $x_0 \circ x_j$ or~$x_i \circ x_j$.
If there is a node computing the last function, you can replace it with a node computing $x_0 \circ x_i$ since the combination~$x_i \circ x_j$ cannot be reused by another tree anyway.
This does not change the total size of the circuit.
So without loss of generality, assume that there are no such nodes.
We choose as a cover all~$i$ for which~$x_0 \circ x_i$ is computed.

Now every edge is covered by one of its endpoints because it corresponds to a tree~$T$ that is computed via an intermediate result.
Hence, the nodes computing Boolean functions on two variables transform into a vertex cover.
There are precisely~$N$ of these.
\end{proof}

\subsection{APX-hardness}\label{subsec:apx_hardness}
We now show how to exploit the reduction to even get APX-hardness of \BSTSO, i.e, to prove \cref{thm:hardness_intro}.

\begin{proposition}\label{pro:vc_to_st_guarantee}
If there exists no~$\alpha$-approximation algorithm for \vc\ in graphs of maximum degree~$\Delta$,
then there does not exist a~$\frac{\Delta + \alpha}{\Delta + 1}$-approximation for \BSTSO\@ either.
\end{proposition}

\begin{proof}
Assume you are given a~$\frac{\Delta + \alpha}{\Delta + 1}$-approximation algorithm~$\mathcal{A}$ for \BSTSO\ and a \vc\ instance in which the maximum degree is at most~$\Delta$.
We once again consider the reduction shown in the proof of \cref{thm:np_hardness} and look at the \BSTSO\ instance obtained by creating one tree per edge.
To solve the vertex cover instance, we solve the \BSTSO\ instance using~$\mathcal{A}$ and take the vertex cover corresponding to this solution.
Let~$\alg_{\text{VC}}$ be the value of the solution computed like this, and let~$\opt_{\text{VC}}$ be the value of the optimum solution to the \vc\ instance.
Let in addition~$\alg_{\text{\BSTSO}}$ be the value of the solution for \BSTSO\ computed by~$\mathcal{A}$ and~$\opt_{\text{\BSTSO}}$ the optimum solution to the \BSTSO\ instance.

We observe that each vertex can cover at most~$\Delta$ edges within any vertex cover.
Hence,
\begin{align}\label{equ:vc_reduction_num_edges}
   \opt_{\text{VC}}\geq \frac{\lvert E \rvert}{\Delta}.
\end{align}
The~$\frac{\Delta + \alpha}{\Delta + 1}$-approximation algorithm~$\mathcal{A}$ for \BSTSO\ implies that, when applied to instances generated from \vc,
\ifbool{long}{
\begin{align*}
   \alg_{\text{VC}} + \lvert E \rvert \eqnoref{=}        & \alg_{\text{\BSTSO}}\\
   ~\eqwithref{$\mathcal{A}$}{\leq}                      & \frac{\Delta + \alpha}{\Delta + 1}  \opt_{\text{\BSTSO}}\\
   \eqnoref{=}                                           & \frac{\Delta + \alpha}{\Delta + 1} (\opt_{\text{VC}} + \lvert E \rvert) \\
   \eqnoref{=}                                           & \frac{\Delta + \alpha}{\Delta + 1}\opt_{\text{VC}} + \left(\frac{\Delta + \alpha}{\Delta + 1} - 1\right)\lvert E \rvert + \lvert E \rvert \\
   ~\eqwithref{\eqref{equ:vc_reduction_num_edges}}{\leq} & \frac{\Delta + \alpha}{\Delta + 1}\opt_{\text{VC}} + \left(\frac{\Delta + \alpha}{\Delta + 1} - 1\right)\Delta\opt_{\text{VC}} + \lvert E \rvert \\
   \eqnoref{=}                                           & \left(\frac{\Delta + \alpha}{\Delta + 1} + \left(\frac{\Delta + \alpha}{\Delta + 1} - 1\right)\Delta\right)\opt_{\text{VC}} + \lvert E \rvert \\
   \eqnoref{=}                                           & \left(\frac{\Delta + \alpha}{\Delta + 1} (\Delta + 1) - \Delta \right)\opt_{\text{VC}} + \lvert E \rvert \\
   \eqnoref{=}                                           & \alpha \opt_{\text{VC}} + \lvert E \rvert.
\end{align*}
}

\ifbool{short}{
   \begin{align*}
      \alg_{\text{VC}} + \lvert E \rvert \eqnoref{=}        & \alg_{\text{\BSTSO}}\\
      ~\eqwithref{$\mathcal{A}$}{\leq}                      & \frac{\Delta + \alpha}{\Delta + 1}  \opt_{\text{\BSTSO}}\\
      \eqnoref{=}                                           & \frac{\Delta + \alpha}{\Delta + 1} (\opt_{\text{VC}} + \lvert E \rvert) \\
      ~\eqwithref{\eqref{equ:vc_reduction_num_edges}}{\leq} & \frac{\Delta + \alpha}{\Delta + 1}\opt_{\text{VC}} + \left(\frac{\Delta + \alpha}{\Delta + 1} - 1\right)\Delta\opt_{\text{VC}} + \lvert E \rvert \\
      \eqnoref{=}                                           & \alpha \opt_{\text{VC}} + \lvert E \rvert.
   \end{align*}
}

So a~$\frac{\Delta + \alpha}{\Delta + 1}$-approximation for \BSTSO\ implies an~$\alpha$-approximation for \vc.
\end{proof}

\begin{proof}[Proof of \cref{thm:hardness_intro}]
As shown by~\cite{min_3_vc_hardness}, it is NP-hard to approximate \vc\ on 4-regular graphs within a factor of~$\frac{53}{52}$.
This implies by \cref{pro:vc_to_st_guarantee} that there does not exist a~$\frac{261}{260}$-approximation for \BSTSO, even for trees with up to three variables.
\end{proof}

\section{Splitting off large intersections for trees with three variables}\label{sec:three_variables}
In this part we want to prove \cref{thm:main_result}~\ref{itm:k_3} from the introduction,
i.e, we want to give an efficient~$\frac{4}{3}$-approximation algorithm for \BSTSO\ on instances with up to three variables.
This algorithm is summarized in \cref{alg:three_variables}.

\begin{algorithm}[htbp]
   \DontPrintSemicolon
   $\T_R = \T$\;
   \While{$\exists S \subsetneq T_1, \dots, T_{\ell} \in \T_R$ such that~$\ell \geq 3$ and~$\lvert S \rvert = 2$}
   {
      Build~$S$ using one vertex.\;
      Build each~$T \in \T_R$ for which~$S \subsetneq T$ from~$S$ using one additional vertex. \;
      $\T_R = \T_R \setminus \{T \in \T_R : S \subsetneq T\}$\;
   }
   Construct the graph~$G = (V(G), E(G))$ with vertex set~$V(G)$ and edge set~$E(G)$ defined as\label{alg:step:three_graph}
   \begin{align*}
      V(G) & \coloneqq \T_R, \\
      E(G) & \coloneqq \left\{\{T, T'\} : \lvert T \cap T' \rvert = 2 \text{ and } T,T' \in \T_R\right\}.\;
   \end{align*}
   Compute a maximum matching~$M$ in~$G$. \label{alg:step:three_matching}\;
   \For{$e=\{T, T'\} \in M$}
   {
      Build~$T \cap T'$ using one vertex.\;
      Build~$T$ and~$T'$ using one additional vertex each.\;
      $\T_R= \T_R \setminus \{T,T'\}$\;
   }
   Build each~$T \in \T_R$ using two vertices.\;

   \caption{\BSTSO\ algorithm for three variables per tree}
   \label{alg:three_variables}
\end{algorithm}

The main idea of this algorithm is that reusing an intermediate result at least three times is already pretty close to optimum.
Thus, our algorithm, formally described as \cref{alg:three_variables}, works as follows: We iteratively take pairs of variables that appear in at least three trees of~$\T$ that have not already been built.
For such a variable pair, let~$\ell \geq 3$ be the number of these trees.
We build all the~$\ell$ trees containing these two variables by adding a node computing the variable pair and~$\ell$ nodes computing the final tree results.
Thus, we need~$\ell + 1$ vertices in total for one such step.
We iterate this step among the remaining trees.

Once there is no variable pair left that is contained in at least three trees, we use a new strategy.
Every pair now occurs in at most two uncovered trees, and we want to find a solution that reuses as many vertices as possible.
This turns out to be equivalent to computing a matching in the graph~$G = (V(G), E(G))$ defined by
\begin{align*}
   V(G) & = \T_R, \\
   E(G) & = \left\{\{T, T'\} : \lvert T \cap T' \rvert = 2; T,T' \in \T_R\right\},
\end{align*}
using~$\T_R$ for the remaining trees.

When you have computed a maximum matching, you can build precisely the pairs of trees corresponding to the matching edges as vertices and use them for two trees each.
The other trees must be built from scratch.
An example for this algorithm can be seen in \cref{fig:three_variables_example}.

\begin{figure}[htbp]
   \begin{center}
      \begin{tikzpicture}
         \newcommand{\y}{1.25}
         \node[scale=1.3] (i1) at (1.5, 3.2){$x_1$};
         \node[scale=1.3] (i2) at (2.5, 3.2){$x_2$};
         \node[scale=1.3] (i3) at (3.5, 3.2){$x_3$};
         \node[scale=1.3] (i4) at (4.5, 3.2){$x_4$};
         \node[scale=1.3] (i5) at (5.5, 3.2){$x_5$};
         \node[scale=1.3] (i6) at (6.5, 3.2){$x_6$};
         \node[scale=1.3] (i7) at (7.5, 3.2){$x_7$};
         \node[scale=1.3] (i8) at (8.5, 3.2){$x_8$};
         \node[scale=1.3] (i9) at (9.5, 3.2){$x_9$};

         \node[] (text) at (5.5, 5) {$\mathcal{T} = \left\{\textcolor{violet}{T_1 = \{1, 2, 3\}, T_2 = \{2, 3, 4\}, T_3 = \{2, 3, 6\},}\right.$};
         \node[] (text2) at (5.5, 4.5) {$ \textcolor{violet}{T_4 = \{3, 5, 7\}, T_5 = \{4, 5, 7\}, T_6 = \{5, 6, 7\}},$};
         \node[] (text3) at (5.5, 4) {$\left. \textcolor{default_blue}{T_7 = \{6, 7, 8\}, T_8 = \{7, 8, 9\}}, \textcolor{mid_green}{T_9 = \{1, 2, 4\}}\right\}$};

         \node[and-gate, fill = violet] (and1) at (3,2) {};
         \draw[->] (i2) -- (and1.input 2);
         \draw[->] (i3) -- (and1.input 1);

         \node[and-gate, fill = violet] (and2) at (6.5,2) {};
         \draw[->] (i5) -- (and2.input 2);
         \draw[->] (i7) -- (and2.input 1);

         \node[and-gate, fill = violet] (and6) at (3.5, 0.5) {};
         \draw[->] (i4) -- (and6.input 1);
         \draw[->] (and1.output) -- (and6.input 2);
         \node[] (t2) at (3.5, 0.5 - \y) {$T_2$};
         \draw[ultra thick, ->] (and6.output) -- (t2);

         \node[and-gate, fill = violet] (and7) at (1.5, 0.5) {};
         \draw[->] (i1) -- (and7.input 2);
         \draw[->] (and1.output) -- (and7.input 1);
         \node[] (t1) at (1.5, 0.5 - \y) {$T_1$};
         \draw[ultra thick, ->] (and7.output) -- (t1);

         \node[and-gate, fill = violet] (and8) at (4.5, 0.5) {};
         \draw[->] (i6) -- (and8.input 1);
         \draw[->] (and1.output) -- (and8.input 2);
         \node[] (t3) at (4.5, 0.5 - \y) {$T_3$};
         \draw[ultra thick, ->] (and8.output) -- (t3);

         \node[and-gate, fill = violet] (and9) at (5.5, 0.5) {};
         \draw[->] (i3) -- (and9.input 2);
         \draw[->] (and2.output) -- (and9.input 1);
         \node[] (t6) at (5.5, 0.5 - \y) {$T_4$};
         \draw[ultra thick, ->] (and9.output) -- (t6);

         \node[and-gate, fill = violet] (and10) at (6.5, 0.5) {};
         \draw[->] (i4) -- (and10.input 2);
         \draw[->] (and2.output) -- (and10.input 1);
         \node[] (t4) at (6.5, 0.5 - \y) {$T_5$};
         \draw[ultra thick, ->] (and10.output) -- (t4);

         \node[and-gate, fill = violet] (and11) at (7.5, 0.5) {};
         \draw[->] (i6) -- (and11.input 1);
         \draw[->] (and2.output) -- (and11.input 2);
         \node[] (t5) at (7.5, 0.5 - \y) {$T_6$};
         \draw[ultra thick, ->] (and11.output) -- (t5);

         \node[and-gate, fill = default_blue] (and3) at (8.5, 2) {};
         \draw[->] (i7) -- (and3.input 2);
         \draw[->] (i8) -- (and3.input 1);

         \node[and-gate, fill = default_blue] (and12) at (8.5, 0.5) {};
         \draw[->] (i6) -- (and12.input 2);
         \draw[->] (and3.output) -- (and12.input 1);
         \node[] (t7) at (8.5, 0.5 - \y) {$T_7$};
         \draw[ultra thick, ->] (and12.output) -- (t7);

         \node[and-gate, fill = default_blue] (and13) at (9.5, 0.5) {};
         \draw[->] (i9) -- (and13.input 1);
         \draw[->] (and3.output) -- (and13.input 2);
         \node[] (t8) at (9.5, 0.5 - \y) {$T_8$};
         \draw[ultra thick, ->] (and13.output) -- (t8);

         \node[and-gate, fill = mid_green] (and4) at (2, 2) {};
         \draw[->] (i1) -- (and4.input 2);
         \draw[->] (i2) -- (and4.input 1);

         \node[and-gate, fill = mid_green] (and5) at (2.5, 0.5) {};
         \draw[->] (i4) -- (and5.input 1);
         \draw[->] (and4.output) -- (and5.input 2);
         \node[] (t9) at (2.5, 0.5 - \y) {$T_9$};
         \draw[ultra thick, ->] (and5.output) -- (t9);
      \end{tikzpicture}
      \caption{Example of \cref{alg:three_variables}: In \textcolor{violet}{purple} we show the gates that arise from building common pairs, in this case~$\{2,3\}$ and~$\{5,7\}$.
      Then we compute a matching on the remaining trees~$T_7$,~$T_8$ and~$T_9$ and find the matching edge corresponding to~$\{7,8\}$. We build those gates in \textcolor{default_blue}{blue}.
      Last, we build all remaining trees (in this case~$T_9$) in \textcolor{mid_green}{green}.}
      \label{fig:three_variables_example}
   \end{center}
\end{figure}
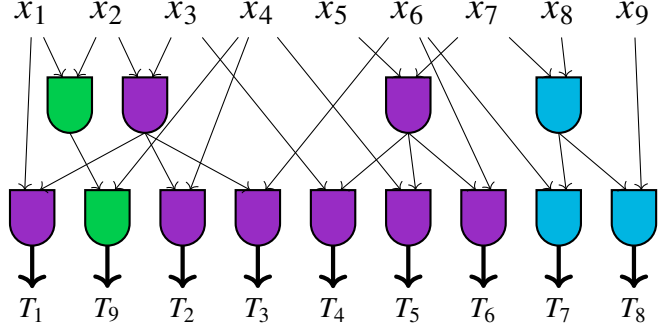
\begin{proof}[Proof of \cref{thm:main_result}~\ref{itm:k_3}]
   If~$\T$ contains trees with two variables, we can build them immediately, just as trees with three variables containing a two-variable tree of~$\T$.
   This is clearly best possible.
   So without loss of generality, assume that each tree in~$\T$ contains three variables.
   We apply \cref{alg:three_variables}, which we analyse in the following.
   For this analysis, we refer to~$\T_R$ as the set of trees remaining after the while-loop, i.e., in \cref{alg:step:three_graph}.
   Look at an optimum solution~$\OPT(\T)$  with value~$\opt(\T)$ for all trees~$\T$.
   It has at least
   \[ \opt(\T) \geq \opt(\T_R) + \lvert \T \backslash \T_R \rvert\]
   gates, where~$\opt(\T_R)$ is the value of an optimum solution for~$\T_R$.
   This is because~$\OPT(\T)$ contains some solution to~$\T_R$ and additionally at least one gate for each tree in~$\T \backslash \T_R$.

   We claim that we solve~$\T_R$ exactly.
   This is due to the fact that every solution to the \BSTSO\ solution of size~$N$ can be transformed into a matching consisting of the reused gates and a set of trees built from scratch.
   This matching has size~$2 \lvert \T_R \rvert - N$.
   In contrast, every matching~$M$ also naturally gives a circuit of size~$2 \lvert \T_R \rvert - \lvert M \rvert$.
   So computing a maximum matching precisely gives a minimum \BSTSO\ solution to~$\T_R$.

   Our algorithm now uses~$\ell + 1$ gates for~$\ell$ trees in every iteration of the while-loop, where~$\ell \geq 3$.
   Thus, the total size of all the trees computed in the first step can be upper bounded by~$\frac 43 \lvert \T \backslash \T_R \rvert$.
   Let~$\alg(\T)$ be the value of the solution of the algorithm.
   Putting everything together, we get

   \begin{align*}
   \frac{\alg(\T)}{\opt(\T)} \leq {}& \frac{\opt(\T_R) + \frac 43 \lvert \T \backslash \T_R \rvert}{\opt(\T_R) + \lvert \T \backslash \T_R \rvert} \leq \frac 43.\qedhere
   \end{align*}
\end{proof}

Instead of constructing the matching in \cref{alg:three_variables} \cref{alg:step:three_matching}, one could also just build all these trees from scratch.
This leads to the same approximation guarantee when analysing it carefully but as we need to do a similar matching approach in \cref{sec:four_variables}, we decided to show it here.

\section{Trees with four variables}\label{sec:four_variables}

\newcommand{\Tsup}{\T_4^{\supsetneq 3}}
\newcommand{\Tmatching}{\T_4^{\operatorname{matching}}}
\newcommand{\Tintersect}{\T_4^{\operatorname{intersect}}}
\newcommand{\Tfourtwo}{\T_4^{\operatorname{depth}2}}
\newcommand{\That}{\T_3'}
\newcommand{\Teasy}{\T^{\operatorname{easy}}}
\newcommand{\Tmain}{\T^{\operatorname{depth}\leq 2}}
Now we want to prove \cref{thm:main_result}~\ref{itm:k_4} from the introduction, i.e, proving that there is a $1.9$-approximation algorithm for instances with trees of up to four variables.
This algorithm is formalized as \cref{alg:four_variables}.

\subsection{Overview of the algorithm}\label{subsec:four_overview}
\begin{algorithm}[htbp]
   \DontPrintSemicolon
   $\Tsup \coloneqq \{ T \in \T_4 : \exists T' \in \T_3 \text{ s.t. } T' \subsetneq T\}$\;
   $\Teasy \leftarrow \Tsup$ \tcp*[r]{trees buildable with one additional gate} \label{alg:step:four_setting_teasy}
   $B \leftarrow \Tsup$ \tcp*[r]{trees where we determined how to build them}
   $\That \leftarrow \T_3$ \tcp*[r]{trees containing three variables}
   \While{$\exists T_1, \dots, T_{\ell} \in \T_4 \backslash B$ s.t.~$\ell \geq 3$ and~$\lvert \bigcap_{i=1}^{\ell} T_i \rvert = 3$ \label{alg:step:four_tintersect_loop}}{
      Build~$S = \bigcap_{i=1}^{\ell} T_i$ with two vertices.\;
      Build each~$\T_i$ using one vertex with input~$S$.\;
      $B = B \cup \{T_1, \dots, T_{\ell}\}$ \tcp*[r]{Trees considered in this loop form~$\Tintersect$.}\label{alg:step:four_end_tintersect_loop}
   }
   Compute a maximum matching~$M^{\max}$ in the graph defined by~\eqref{equ:four_matching_graph}, i.e.,
   \begin{align*}
   V(G) & = \T_4 \setminus B, \,E(G) = \{ \{ T, T' \} : \lvert T \cap T' \rvert = 3 \}.
   \end{align*}\label{alg:step:four_max_matching}\;
   \vspace{-\baselineskip}
   \For{$\{T, T'\} \in M^{\max}$ \label{alg:step:four_tmatching_loop}}{
      $\That = \That \cup \left(T \cap T'\right)$\;
      $\Teasy = \Teasy \cup \{T, T'\}$ \tcp*[r]{Trees considered in this loop form~$\Tmatching$.}\label{alg:step:four_end_tmatching_loop}
   }
   $\Tmain \coloneqq \T_2 \cup \That \cup (\T_4 \setminus B$) \tcp*[r]{Trees in~$\T_4 \setminus B$ are called~$\Tfourtwo$}
   Construct a hypergraph~$H = (V(H), E(H))$ with \label{alg:step:four_graph_construction}
   \begin{align*}
      V(H) = {} & \{U : U \subseteq T \in \Tmain, \lvert U \rvert = 2\} \\
      E(H) = {} & \left\{\{U_{\alpha(1)}^T, U_{\alpha(2)}^T, U_{\alpha(3)}^T\} : T \in \T_4 \setminus B, \alpha(i) \in \{i, \bar{i}\}\right\} \\
      & \cup \left\{\{U \subsetneq T\} : \lvert U \rvert = 2, T \in \That \right\}\\
      & \cup \left\{T : T \in \T_2\right\}
   \end{align*}
   where the~$U_{\alpha(i)}$ are the two element subsets of~$T$, i.e., for~$T = \{1, 2, 3, 4\}$ we have
   \begin{align*}
      & U_{1}^T = \{1, 2\},       & U_{2}^T = \{1, 3\},        & & U_{3}^T = \{1, 4\}, \\
      & U_{\bar{1}}^T = \{3, 4\}, & U_{\bar{2}}^T = \{2, 4\},  & & U_{\bar{3}}^T = \{2, 3\}.
   \end{align*}\;
   \vspace{-\baselineskip}
   Compute a vertex cover~$C$ in~$H$ as the union of all vertices of a maximal matching. \label{alg:step:four_solve_vc}\;
   Remove unnecessary vertices from~$C$. \label{alg:step:four_remove_vertices}\;
   \If{$\lvert C \rvert > \lvert \T_2 \cup \T_3' \rvert + 2 \lvert \Tfourtwo \rvert$}{
      Build each tree~$T \in \Tmain$ separately. \label{alg:step:four_variables_alternative_solution} \;
   }
   \Else{
      Build a node for each~$U$ with~$v_U \in C$. \label{alg:step:vc_to_bstso_trafo}\;
      Build each~$T \in \Tmain \setminus \T_2$ by combining trees for~$U$ and~$T \setminus U$ with one node.\;
   }
   $B = B \cup \Tmain$\;
   \ForEach{$T \in \Teasy$}{
   Build a tree for~$T$ using one vertex from a subtree of~$T$ that was built previously. \label{alg:step:building_teasy}\;
      $B = B \cup \{T\}$\;
   }

   \caption{\BSTSO\ algorithm for up to four variables per tree}
   \label{alg:four_variables}
\end{algorithm}

Our approach for the algorithm is as follows (compare \cref{alg:four_variables}).
In several steps, we reduce the instance until the remaining, although slightly modified instance can be built with depth~$2$ optimally, i.e., such that the longest path in one optimum circuit is of length 2.
Note that we do not optimize the depth here but rather we are interested in this reduction because the remaining instance has a special structure and allows for a better algorithm.
Thus, we first get rid of all trees not suitable for this approach.
It will turn out that the trees that need to be removed in this reduction are rather cheap to build anyway.
Regarding notation, let~$\T_i \subseteq \T$ be the set of those trees that contain exactly~$i$ variables for~$i \in \{2, 3, 4\}$.

The reduction steps are the following ones.
First, we look at all the trees~$T \in \T$ of size~$4$ that contain a tree~$T' \in \T$ of size~$3$.
These trees can be built easily in the end by reusing the result of tree~$T'$ and adding a single vertex to compute~$T$.
We call these trees~$\Tsup$.
We add them to a set~$\Teasy$ of vertices that we will build in the end (see \cref{alg:four_variables} \cref{alg:step:four_setting_teasy}).

Next, we need to do something similar to the case of three variables per tree in \cref{sec:three_variables}.
If at least~$\ell \geq 3$ trees with four variables intersect in three variables, we build them immediately (see~\crefrange{alg:step:four_tintersect_loop}{alg:step:four_end_tintersect_loop}).
The set of trees built in this step is called~$\Tintersect$.

Now remaining are~$\T_2, \T_3$ and a part of~$\T_4$ for which each triple of variables appears in at most two trees.
Once again, we use a similar approach as in \cref{sec:three_variables}.
We create a graph~$G = (V(G), E(G))$ defined by
\begin{align}\label{equ:four_matching_graph}
V(G) & = \T_4 \setminus (\Tintersect \cup \Tsup), E(G) = \{ \{ T, T' \} : \lvert T \cap T' \rvert = 3 \}.
\end{align}
In this graph, we can compute a maximum matching to decide which pairs of four-variable trees shall be built sharing three variables (see \cref{alg:step:four_max_matching}).
However, we need to be very careful here.
Instead of immediately building the needed five vertices for the two trees, we add their intersection to the remaining instance.
More precisely, for each matching edge~$\{T,T'\}$, we add~$T \cap T'$ to a new set~$\T_3'$.
We in addition put all trees from~$\T_3$ into~$\T_3'$, too (see~\crefrange{alg:step:four_tmatching_loop}{alg:step:four_end_tmatching_loop}).
So~$\T_3'$ will be the new set of trees of size 3 that we want to build.
Let the set of trees covered by the matching be~$\Tmatching$.
Then we add~$\Tmatching$ to~$\Teasy$, i.e., to the trees to be built in the end.
This makes sense as building~$\Tmatching$ from $\T_3'$ is indeed easy.

We now solve the remaining instance including the modification, i.e., building trees for all of~$\T_3'$, too.
This is the last part of our algorithm.
No pair of remaining trees shares three variables and this implies that the trees can be built with depth  (at most)~$2$ as will be shown in \cref{lem:opt_2_4_is_nice}.
We call the remaining trees~$\Tmain$ and set~$\Tfourtwo \coloneqq \Tmain \cap \T_4$.

For~$\Tmain$ we now use a vertex cover model within a hypergraph defined as follows (see \cref{alg:step:four_graph_construction}).
\begin{align}\label{equ:four_hypergraph_definition}
   V(H) \coloneqq & \{U : U \subseteq T \in \Tmain, \lvert U \rvert = 2\} \nonumber\\
   E(H) \coloneqq & \left\{\{U_{\alpha(1)}^T, U_{\alpha(2)}^T, U_{\alpha(3)}^T\} : T \in \Tfourtwo, \alpha(i) \in \{i, \bar{i}\}\right\} \\
   & \cup \left\{\{U \subsetneq T\} : \lvert U \rvert = 2, T \in \That \right\} \nonumber\\
   & \cup \left\{T : T \in \T_2\right\} \nonumber
\end{align}
where the~$U_{\alpha(i)}$ are the two element subsets of~$T$, i.e., for~$T = \{1, 2, 3, 4\}$ we have
\begin{align*}
   & U_{1}^T = \{1, 2\},       & U_{2}^T = \{1, 3\},        & & U_{3}^T = \{1, 4\}, \\
   & U_{\bar{1}}^T = \{3, 4\}, & U_{\bar{2}}^T = \{2, 4\},  & & U_{\bar{3}}^T = \{2, 3\}.
\end{align*}
We want to build a gate computing~$\{x_i,x_j\}$ if and only if vertex~$\{i, j\}$ of the hypergraph is chosen by the cover.
By the definition above we ensure the following properties.
\begin{itemize}
   \item Each vertex corresponding to a tree in~$\T_2$ is chosen in the vertex cover.
   \item For each tree~$T \in \T_3'$ at least one of the two-element subsets is chosen by the vertex cover.
   So we need at most one additional gate per~$T \in \T_3'$ to build the tree compared to the vertex cover.
   \item For each tree~$T \in \Tfourtwo$ there exists an~$i$ such that~$U_{i}^T$ and~$U_{\bar{i}}^T$ are chosen.
   Thus,~$T$ can also be built with one additional gate compared to the vertex cover.
\end{itemize}
Thus, we know how to get the circuit we are looking for from a computed vertex cover (compare \cref{alg:step:vc_to_bstso_trafo}).
It remains to solve the vertex cover instance.
For the vertex cover, we compute a maximal matching and add all vertices contained in a matching edge (see \cref{alg:step:four_solve_vc}).
If this vertex cover is more expensive than just taking one vertex per tree in~$\Tmain$ plus one per tree in~$\Tfourtwo$, we build~$\Tmain$ from scratch (see \cref{alg:step:four_variables_alternative_solution}).

In the end, we conclude the algorithm by building all the trees we have put into~$\Teasy$ with a single gate from already built trees (see \cref{alg:step:building_teasy}).

\begin{figure}[htbp]
   \begin{center}
      \begin{tikzpicture}[scale = 1.0]
         \node[scale=1.3] (i1) at (1.5, 3.2){$x_1$};
         \node[scale=1.3] (i2) at (2.5, 3.2){$x_2$};
         \node[scale=1.3] (i3) at (3.5, 3.2){$x_3$};
         \node[scale=1.3] (i4) at (4.5, 3.2){$x_4$};
         \node[scale=1.3] (i5) at (5.5, 3.2){$x_5$};
         \node[scale=1.3] (i6) at (6.5, 3.2){$x_6$};
         \node[scale=1.3] (i7) at (7.5, 3.2){$x_7$};
         \node[scale=1.3] (i8) at (8.5, 3.2){$x_8$};

         \node[and-gate] (and1) at (3,2) {};
         \node[and-gate] (and2) at (2.5, 1) {};
         \node[] (output1) at (1.5, -1) {$\in \T_3$};

         \draw[->] (i2) -- (and1.input 2);
         \draw[->] (i3) -- (and1.input 1);
         \draw[->] (and1.output) -- (and2.input 1);
         \draw[->] (i1) -- (and2.input 2);
         \draw[ultra thick, ->] (and2.output) -- (output1);

         \node[and-gate] (and3) at (5,2) {};
         \node[and-gate] (and5) at (4,0) {};
         \node[] (output2) at (3.5, -2) {$\in \Tfourtwo$};

         \draw[->] (i4) -- (and3.input 2);
         \draw[->] (i5) -- (and3.input 1);
         \draw[->] (and1.output) -- (and5.input 2);
         \draw[->] (and3.output) -- (and5.input 1);
         \draw[ultra thick, ->] (and5.output) -- (output2);

         \node[and-gate] (and6) at (7,2) {};
         \node[and-gate] (and7) at (6,1) {};
         \node[and-gate] (and8) at (5.5,0) {};
         \node[] (output3) at (5.5, -2) {$\in \Tmatching$};

         \draw[->] (i6) -- (and6.input 2);
         \draw[->] (i7) -- (and6.input 1);
         \draw[->] (i5) -- (and7.input 2);
         \draw[->] (and6.output) -- (and7.input 1);
         \draw[->] (and7.output) -- (and8.input 1);
         \draw[->] (i3) -- (and8.input 2);
         \draw[ultra thick, ->] (and8.output) -- (output3);

         \node[and-gate] (and9) at (6.5, 0) {};
         \node[] (output4) at (8, -2) {$\in \Tmatching$};

         \draw[->] (i8) -- (and9.input 1);
         \draw[->] (and7.output) -- (and9.input 2);
         \draw[ultra thick, ->] (and9.output) -- (output4);

         \node[and-gate] (and10) at (3, 0) {};
         \node[] (output5) at (2, -2) {$\in \Tsup$};

         \draw[->] (and2.output) -- (and10.input 2);
         \draw[->] (i4) -- (and10.input 1);
         \draw[ultra thick, ->] (and10.output) -- (output5);

         \node[] (output6) at (8.5, 0) {$\in \T_2$};
         \draw[ultra thick, ->] (and6.output) -- (output6);

      \end{tikzpicture}
      \caption{Example partition of the instance as given by~\eqref{equ:bstso_four_partition}.}
      \label{fig:four_variables_partition}
   \end{center}
\end{figure}
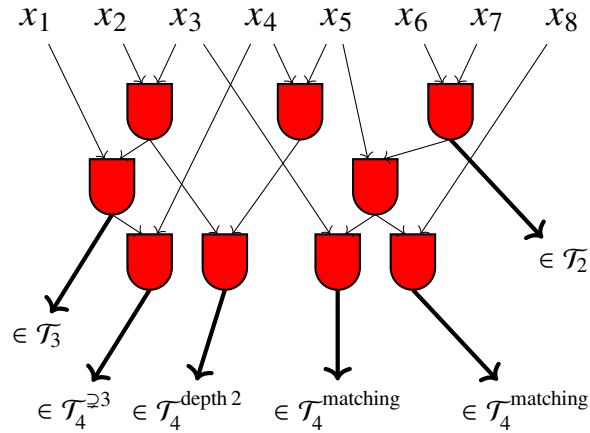

So now that we have defined all the subinstances arising in the different steps, we can write
\begin{align}\label{equ:bstso_four_partition}
   \T = \T_2 \discup \T_3 \discup \Tsup \discup \Tintersect \discup \Tmatching  \discup \Tfourtwo.
\end{align}
An example of how this subdivision can look like is given by \cref{fig:four_variables_partition}.
Using this description of the algorithm we now prove the desired approximation guarantee.
For this, we are going over the algorithm again, proving the some quantitative results for the different steps.

\subsection{Trees containing trees of size 3}\label{subsec:four_variables_t_sup}
In order not to lose too much size on trees that are easy to build, we need to be very careful when building the
trees that share a lot of variables with each other.
First, it is easy to see that trees in~$\T_4$ that contain a tree of~$\T_3$ need exactly one additional gate compared to a solution of the remaining instance.
We build this gate just after building the remaining instance.

\begin{lemma}\label{lem:removing_t_sup}
   We have
   \begin{align}\label{equ:remove_tsup}
      \opt(\T) = \opt\left(\T \setminus \Tsup\right) + \lvert \Tsup \rvert.
   \end{align}
\end{lemma}
\begin{proof}
   On the one hand, the vertex computing the final result for a~$T \in \Tsup$ cannot be reused in any tree~$T' \in \T \setminus \Tsup$, especially not in an optimum one.
   This implies~$``\geq``$.

   On the other hand, you can make every solution for~$\T \setminus \Tsup$ a solution for~$\T$ by adding one vertex per tree in~$\Tsup$ that computes the final result from a tree with three inputs and the fourth input.
\end{proof}

\subsection{Trees with large intersection}\label{subsec:four_variables_large_intersection}

For trees with four inputs that share at least three variables with other four input trees we use the following procedure.
First, if there is a 3-tuple of variables that is contained in at least three four-input trees,
we build a circuit for those immediately, compare \cref{alg:four_variables}~\crefrange{alg:step:four_tintersect_loop}{alg:step:four_end_tintersect_loop}.
The 3-tuple can be computed using two gates and for each four-input tree, we need one additional gate.
So we need~$k+2$ gates to build~$k$ trees where~$k \geq 3$.
These trees are now considered built and no longer available for further tuples.
We choose such 3-tuples greedily as long as there exist some.
All trees chosen within within this step are called~$\Tintersect$.

\begin{lemma}\label{lem:four_large_intersection}
   We have
   \begin{align}\label{equ:remove_intersection}
      \opt\left(\T \setminus \Tsup\right) \geq \opt\left(\T \setminus (\Tsup \cup \Tintersect)\right) + \lvert \Tintersect \rvert.
   \end{align}
\end{lemma}
\begin{proof}
   The final gate computing a~$T \in \Tintersect$ can never be reused.
   So removing these gates from the optimum solution for~$\T \setminus \Tsup$ still yields a solution for the remaining instance.
\end{proof}

Now that we got rid of all 3-tuples arising more than twice, we have to be careful on the remaining part.
Notice that due to \cref{lem:four_large_intersection}, it will be enough to compare to an optimum solution for the remaining tree set.
This is important since we have no idea how the optimum solution for the original tree set~$\T$ might look like.
We now build pairs of trees that share three variables, compare \cref{alg:four_variables}~\crefrange{alg:step:four_tmatching_loop}{alg:step:four_end_tmatching_loop}.
Those pairs are determined by a maximum matching algorithm to make sure that we actually get the maximum number of pairs (see \cref{alg:step:four_max_matching}).

The matching algorithm works on the graph~$G$ defined by~\eqref{equ:four_matching_graph}.
Here, each edge~$e = \{T, T'\}$ is related to a three-input set~$S(e) \coloneqq T \cap T'$.

We call the set of all vertices in these trees~$\Tmatching$ and we add all 3-tuples we obtain from these pairs together with~$\T_3$ to a new set~$\That$.

\begin{lemma}\label{lem:four_matching}
   Let~$\That$ be the union of~$\T_3$ and all 3-tuples that appear as the intersection of a pair in~$\Tmatching$.
   Let~$1 \leq \alpha \leq 2$.
   Assume you are given an~$\alpha$-approximation algorithm~$\mathcal{A}$ for~$\Tmain \coloneqq \T_2 \cup \That \cup \Tfourtwo$.
   Then there exists a~$(1 + \alpha / 2)$-approximation for~$\mathcal{S} \coloneqq \T_2 \cup \T_3 \cup \Tfourtwo \cup \Tmatching$.
\end{lemma}
\begin{proof}
   Let~$\OPT(\mathcal{S})$ be an optimum solution for~$\mathcal{S}$ that contains a minimum number of vertices computing functions of exactly three variables.

   Our algorithm computes a maximum matching~$M^{\max}$ in~$G$ as defined by~\eqref{equ:four_matching_graph}, applies algorithm~$\mathcal{A}$ to the computed instance and
   then adds the necessary additional vertices to the computed solution in order to make it a solution for~$\mathcal{S}$ (see \cref{fig:matching_flow}).

   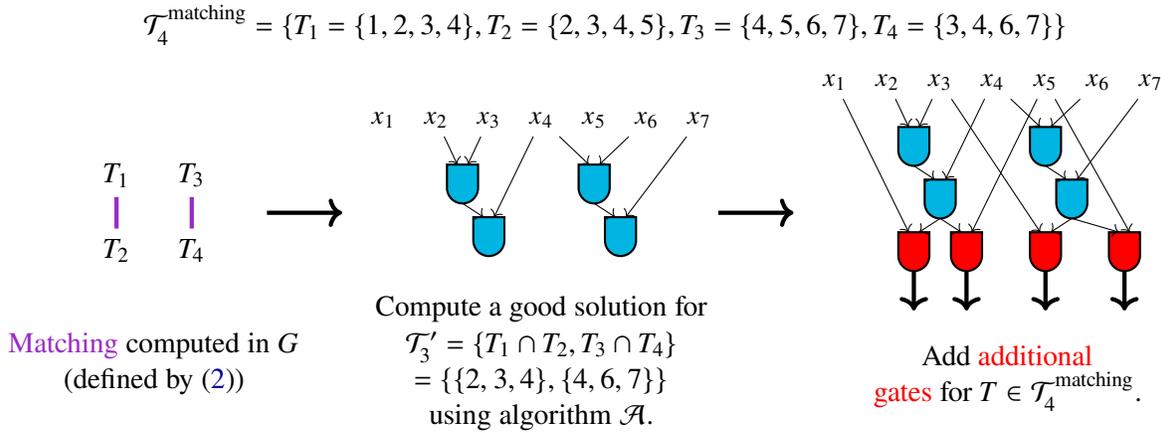
\begin{figure}[htbp]
      \begin{center}
         \begin{tikzpicture}[scale = 1.0]
            \begin{scope}[shift={(1, 2.5)}, scale=1.0, every node/.append style={transform shape}]
               \node at (0,0) (instance) {$\Tmatching = \{T_1 = \{1, 2, 3, 4\}, T_2 = \{2,3,4,5\}, T_3 = \{4, 5, 6, 7\}, T_4 = \{3, 4, 6, 7\}\}$};
            \end{scope}
            \begin{scope}[shift={(-5, 0)}, scale=1.0, every node/.append style={transform shape}]
               \node at (-0.5, 0.5) (t1) {$T_1$};
               \node at (-0.5, -0.5) (t2) {$T_2$};
               \node at (0.5, 0.5) (t3) {$T_3$};
               \node at (0.5, -0.5) (t4) {$T_4$};

               \draw[color = violet, ultra thick] (t1) -- (t2);
               \draw[color = violet, ultra thick] (t3) -- (t4);

               \node at (0, -2) (text) [align=center] {\textcolor{violet}{Matching} computed in~$G$ \\ (defined by \eqref{equ:four_matching_graph})};
            \end{scope}
            \draw[ultra thick, ->] (-3.5, 0) -- (-2.5, 0);
            \begin{scope}[shift={(-3, -1)}, scale=0.7, every node/.append style={transform shape}]
               \node[scale=1.3] (i1) at (1.5, 3.2){$x_1$};
               \node[scale=1.3] (i2) at (2.5, 3.2){$x_2$};
               \node[scale=1.3] (i3) at (3.5, 3.2){$x_3$};
               \node[scale=1.3] (i4) at (4.5, 3.2){$x_4$};
               \node[scale=1.3] (i5) at (5.5, 3.2){$x_5$};
               \node[scale=1.3] (i6) at (6.5, 3.2){$x_6$};
               \node[scale=1.3] (i7) at (7.5, 3.2){$x_7$};

               \node[and-gate, fill=default_blue] (and1) at (3, 2) {};
               \node[and-gate, fill=default_blue] (and2) at (3.5, 1) {};
               \node[and-gate, fill=default_blue] (and3) at (5.5, 2) {};
               \node[and-gate, fill=default_blue] (and4) at (6, 1) {};

               \draw[->] (i2) -- (and1.input 2);
               \draw[->] (i3) -- (and1.input 1);
               \draw[->] (i4) -- (and3.input 2);
               \draw[->] (i6) -- (and3.input 1);
               \draw[->] (i4) -- (and2.input 1);
               \draw[->] (i7) -- (and4.input 1);
               \draw[->] (and1.output) -- (and2.input 2);
               \draw[->] (and3.output) -- (and4.input 2);

            \end{scope}
            \node at (-3 + 0.7 * 4.5, -2) (text2) [align=center] {Compute a good solution for \\ $\That= \{T_1 \cap T_2, T_3 \cap T_4\}$ \\ $= \{\{2, 3, 4\}, \{4, 6, 7\}\}$ \\ using algorithm~$\mathcal{A}$.};
            \draw[ultra thick, ->] (2.5, 0) -- (3.5,0);
            \begin{scope}[shift={(3, -0.5)}, scale=0.7, every node/.append style={transform shape}]
               \node[scale=1.3] (i1) at (1.5, 3.2){$x_1$};
               \node[scale=1.3] (i2) at (2.5, 3.2){$x_2$};
               \node[scale=1.3] (i3) at (3.5, 3.2){$x_3$};
               \node[scale=1.3] (i4) at (4.5, 3.2){$x_4$};
               \node[scale=1.3] (i5) at (5.5, 3.2){$x_5$};
               \node[scale=1.3] (i6) at (6.5, 3.2){$x_6$};
               \node[scale=1.3] (i7) at (7.5, 3.2){$x_7$};

               \node[and-gate, fill=default_blue] (and1) at (3, 2) {};
               \node[and-gate, fill=default_blue] (and2) at (3.5, 1) {};
               \node[and-gate, fill=default_blue] (and3) at (5.5, 2) {};
               \node[and-gate, fill=default_blue] (and4) at (6, 1) {};

               \draw[->] (i2) -- (and1.input 2);
               \draw[->] (i3) -- (and1.input 1);
               \draw[->] (i4) -- (and3.input 2);
               \draw[->] (i6) -- (and3.input 1);
               \draw[->] (i4) -- (and2.input 1);
               \draw[->] (i7) -- (and4.input 1);
               \draw[->] (and1.output) -- (and2.input 2);
               \draw[->] (and3.output) -- (and4.input 2);

               \node[and-gate] (and5) at (3, 0) {};
               \node[and-gate] (and6) at (4, 0) {};
               \node[and-gate] (and7) at (7, 0) {};
               \node[and-gate] (and8) at (5.5, 0) {};

               \draw[->] (and2.output) -- (and5.input 1);
               \draw[->] (i1) -- (and5.input 2);
               \draw[->] (and2.output) -- (and6.input 2);
               \draw[->] (i5) -- (and6.input 1);
               \draw[->] (and4.output) -- (and7.input 2);
               \draw[->] (i5) -- (and7.input 1);
               \draw[->] (i3) -- (and8.input 2);
               \draw[->] (and4.output) -- (and8.input 1);
               \draw[->, ultra thick] (and5.output) -- ($(and5.output) + (0, -0.75)$);
               \draw[->, ultra thick] (and6.output) -- ($(and6.output) + (0, -0.75)$);
               \draw[->, ultra thick] (and7.output) -- ($(and7.output) + (0, -0.75)$);
               \draw[->, ultra thick] (and8.output) -- ($(and8.output) + (0, -0.75)$);
            \end{scope}
            \node at (3 + 0.7 * 4.5 + 0.2, -2.2) (text2) [align=center] {Add \textcolor{mid_red}{additional} \\ \textcolor{mid_red}{gates} for~$T \in \Tmatching$.};
         \end{tikzpicture}
         \caption{Example for the algorithm used to prove \cref{lem:four_matching}.~$\T_3$ is empty in this case,~$\Tmatching$ is shown at the top.}
         \label{fig:matching_flow}
      \end{center}
   \end{figure}

   Looking at~$\OPT(\mathcal{S})$, there might exist some pairs of trees~$T, T'$ whose final results are build from a common three-variable vertex~$T \cap T'$.
   For each of those pairs, there exists a corresponding edge in~$G$ (defined by~\eqref{equ:four_matching_graph}).
   All these edges form a matching~$M^{\opt}$ in~$G$.
   Let the computed maximum matching in~$G$ be~$M^{\max}$.
   We consider the symmetric difference~$M^{\max} \symdiff M^{\opt}$, i.e., the set of edges that is either in~$M^{\max}$ or in~$M^{\opt}$ (but not in both).

   Since~$M^{\max}$ is maximum,~$M^{\max} \symdiff M^{\opt}$ consists of alternating paths and cycles, none of which is an~$M^{\max}$ augmenting path.
   So each path starts or ends with an edge contained in~$M^{\max}$.
   Thus, the number of inner edges belonging to~$M^{\max}$ is at most the number of edges in~$M^{\opt}$.

   For each edge~$e \in M^{\max}$ there are two cases:
   Either there exists a vertex~$t$ in the optimum solution~$\OPT(\mathcal{S})$ that computes a two-input subset of the three-input set~$S(e)$ (compare \cref{fig:matching_difference-1}).
   Then it is possible to build both trees~$T, T' \in e$ from~$t$ using a total of~$3$ gates: One to compute the three-input set~$S(e)$ and one for each~$T$ and~$T'$.

   \begin{figure}[htbp]
      \begin{center}
         \begin{subfigure}{0.45\textwidth}
            \begin{center}
            \begin{tikzpicture}[scale = 1.0]
               \node[scale=1.3] (i1) at (1.5, 3.2){$x_1$};
               \node[scale=1.3] (i2) at (2.5, 3.2){$x_2$};
               \node[scale=1.3] (i3) at (3.5, 3.2){$x_3$};
               \node[scale=1.3] (i4) at (4.5, 3.2){$x_4$};
               \node[scale=1.3] (i5) at (5.5, 3.2){$x_5$};

               \node[and-gate, logic gate inputs=nnn, fill=default_blue] (and1) at (3.5,1.2) {};
               \node[and-gate] (and2) at (2.75, -1) {};
               \node[] (output1) at (2.75, -2.25) {$T_1$};

               \node[and-gate] (and3) at (4.25, -1) {};
               \node[] (output2) at (4.25, -2.25) {$T_2$};

               \node[and-gate, color = black!35, fill = or_green!50, logic gate inputs=nn] (and_t1) at (2.5,2) {\rot{$t$}};
               \node[and-gate, color = black!35, fill = or_green!50, logic gate inputs=nn] (and_t2) at (1.7,1) {};

               \draw[->, draw = black!35] (i2) -- (and_t1.input 2);
               \draw[->, draw = black!35] (i3) -- (and_t1.input 1);
               \draw[->, draw = black!35] (i1) -- (and_t2.input 2);
               \draw[->, draw = black!35] (and_t1.output) -- (and_t2.input 1);

               \draw[->, draw = black!35] (and_t1.output) -- (and2.input 2);
               \draw[->, draw = black!35] (and_t2.output) -- (and2.input 2);

               \draw[->, draw = black!35] (i4) [out=-90, in=30] to (and2.input 1);

               \draw[->, dotted] (i3) -- (and1.input 2);
               \draw[->, dotted] (i4) -- (and1.input 1);
               \draw[->, dotted] (i2) -- (and1.input 3);
               \draw[->] (and1.output) -- (and2.input 1);
               \draw[->] (i1) -- (and2.input 2);
               \draw[ultra thick, ->] (and2.output) -- (output1);

               \draw[->] (i5) -- (and3.input 1);
               \draw[->] (and1.output) -- (and3.input 2);
               \draw[ultra thick, ->] (and3.output) -- (output2);

               \node[and-gate, color = black!35, fill = or_green!50, label=below:{ \small $\Opt(\T)$}, scale=0.5] at (1,0) (optlib) {};
               \node[and-gate, fill = default_blue, label=270:{\small $\That$}, scale=0.5] at (1, -1) (optlibprime) {};

            \end{tikzpicture}
            \caption{Sketch of the first case, i.e, one of~$T_1$ and~$T_2$ is built in the optimum solution using a gate that computes an intermediate result for~$T_1 \cap T_2$.
            In this example,~$T_1$ is built using a gate for~$x_2 \land x_3$.
            Now the \textcolor{default_blue}{blue gate} could take as inputs~$t$ computing~$(x_2 \land x_3)$ and~$x_4$.}
            \label{fig:matching_difference-1}
            \end{center}
         \end{subfigure}
         \hfill
         \begin{subfigure}{0.45 \textwidth}
            \begin{center}
            \begin{tikzpicture}[scale = 1.0]
               \node[scale=1.3] (i1) at (1.5, 3.2){$x_1$};
               \node[scale=1.3] (i2) at (2.5, 3.2){$x_2$};
               \node[scale=1.3] (i3) at (3.5, 3.2){$x_3$};
               \node[scale=1.3] (i4) at (4.5, 3.2){$x_4$};
               \node[scale=1.3] (i5) at (5.5, 3.2){$x_5$};

               \node[and-gate, logic gate inputs=nnn, fill=default_blue] (and1) at (3.5,1.2) {};
               \node[and-gate] (and2) at (2.75, -1) {};
               \node[] (output1) at (2.75, -2.25) {$T_1$};

               \node[and-gate] (and3) at (4.25, -1) {};
               \node[] (output2) at (4.25, -2.25) {$T_2$};

               \node[and-gate, color = black!35, fill = or_green!50, logic gate inputs=nn] (and_t1) at (2,2) {};
               \node[and-gate, color = black!35, fill = or_green!50, logic gate inputs=nn] (and_t2) at (2.5,1) {};

               \draw[->, draw = black!35] (i1) -- (and_t1.input 2);
               \draw[->, draw = black!35] (i2) -- (and_t1.input 1);

               \draw[->, draw = black!35] (and_t1.output) -- (and_t2.input 2);
               \draw[->, draw = black!35] (i3) -- (and_t2.input 1);

               \draw[->, draw = black!35] (and_t2.output) -- (and2.input 2);

               \node[and-gate, color = black!35, fill = or_green!50, logic gate inputs=nn] (and_t3) at (5,2) {};
               \node[and-gate, color = black!35, fill = or_green!50, logic gate inputs=nn] (and_t4) at (4.5,1) {};

               \draw[->, draw = black!35] (i4) -- (and_t3.input 2);
               \draw[->, draw = black!35] (i5) -- (and_t3.input 1);

               \draw[->, draw = black!35] (and_t3.output) -- (and_t4.input 1);
               \draw[->, draw = black!35] (i3) -- (and_t4.input 2);

               \draw[->, draw = black!35] (and_t4.output) -- (and3.input 1);

               \draw[->, draw = black!35] (i4) [out=-100, in=40] to (and2.input 1);
               \draw[->, draw = black!35] (i2) [out=-80, in=140] to (and3.input 2);

               \draw[->, dotted] (i3) -- (and1.input 2);
               \draw[->, dotted] (i4) -- (and1.input 1);
               \draw[->, dotted] (i2) -- (and1.input 3);
               \draw[->] (and1.output) -- (and2.input 1);
               \draw[->] (i1) [out=-90, in=150] to  (and2.input 2);
               \draw[ultra thick, ->] (and2.output) -- (output1);

               \draw[->] (i5) [out=-90, in=30] to (and3.input 1);
               \draw[->] (and1.output) -- (and3.input 2);
               \draw[ultra thick, ->] (and3.output) -- (output2);

               \node[and-gate, color = black!35, fill = or_green!50, label=below:{ \small $\Opt(\T)$}, scale=0.5] at (1,0) (optlib) {};
               \node[and-gate, fill=default_blue, label=270:{\small $\That$}, scale=0.5] at (1, -1) (optlibprime) {};

            \end{tikzpicture}
            \caption{Sketch of the second case, i.e., both~$T_1$ and~$T_2$ are built from gates computing a function on three variables in the optimum solution.
            Both these gates do not use a reusable gate for~$S(e)$.
            An additional gate as intermediate result for the \textcolor{cyan}{blue gate} is needed.
            This cannot happen too often.}
            \label{fig:matching_difference-2}
            \end{center}
         \end{subfigure}
         \hfill
         \caption{Sketches of both cases distinguished in the proof of \cref{lem:four_matching}. In \textcolor{mid_red}{red}, we can see the final gates that both the optimum and our solution for~$\T$ contain.
         In \textcolor{or_green!50}{green} you can see the other gates of the (partial) optimum solution.
         In \textcolor{default_blue}{blue}, you can see the tree of~$\That$ that our algorithm wants to build instead.}
         \label{fig:matching_difference}
      \end{center}
   \end{figure}

   In some other cases, there might not be such a~$t$ (compare \cref{fig:matching_difference-2}).
   This can only happen if neither of~$T$ or~$T'$ is build with depth~$2$ in~$\OPT(\mathcal{S})$ because one of the two sets that a four-input tree is built from must be contained in~$T \cap T'$.
   In addition, the three-input functions that~$T$ and~$T'$ are built from in~$\OPT(\mathcal{S})$ must be different ones.
   Each of them must be reused by another four-input tree, otherwise we could rebuild the tree for~$T$ or~$T'$ as a depth~$2$ tree, contradicting our choice of~$\OPT(\mathcal{S})$.
   So speaking in terms of~$G$, being in the second case means that~$e$ must be an inner edge of a path or cycle in~$M^{\max} \symdiff M^{\opt}$.
   So the second case cannot occur more than~$\lvert M^{\opt} \rvert$ times.

   We now try to get a relation between the optimum values for~$\mathcal{S}$ and~$\Tmain$.
   Let us start with~$\OPT(\mathcal{S})$.
   We now imagine to remove all final vertices computing trees~$T \in \Tmatching = V(M^{\max})$ and all three-input vertices corresponding to edges of~$M^{\opt}$.
   Then we are left with a solution for~$\T_2 \cup \T_3 \cup \Tfourtwo$.
   Now we want to add vertices to compute the trees~$\That \setminus \T_3$, i.e., $S(e)$ for each~$e \in M^{\max}$.
   For each edge~$e$ of the first kind, it suffices to add one vertex to compute~$S(e)$.
   For each edge~$e$ of the second kind, we need to add up to two vertices.
   This means, we add a total of at most~$\lvert M^{\max} \rvert + \lvert M^{\opt} \rvert$ vertices because only inner edges of~$M^{\max} \symdiff M^{\opt}$ can be of the second kind.
   After this procedure, we have a solution to~$\Tmain$, which is at least as large as the optimum one.
   This implies

   \begin{align}\label{equ:four_matching_comparison}
   \opt\left( \mathcal{S} \right) - 2 \lvert M^{\max} \rvert - \lvert M^{\opt} \rvert + \lvert M^{\max} \rvert + \lvert M^{\opt} \rvert & \notag\\
   = \opt\left( \mathcal{S} \right) -\lvert M^{\max} \rvert & \geq  \opt\left(\Tmain \right).
   \end{align}
   What is more, it takes exactly~$2 \lvert M^{\max} \rvert$ gates to complete a solution for~$\Tmain$ to a solution for~$\mathcal{S}$.
   Using additionally that
   \begin{align}\label{equ:max_leq_s}
      \lvert M^{\max} \rvert = \frac 12 \lvert \Tmatching \rvert \leq \frac 12 \opt(\mathcal{S})
   \end{align}
   we get
   \begin{align*}
   \alg\left(\mathcal{S}\right)
   \eqnoref{=}                                                    & \alg\left( \Tmain \right) + 2 \lvert M^{\max} \rvert \\
   \eqnoref{\leq}                                                 & \alpha \opt\left(\Tmain \right) + 2 \lvert M^{\max} \rvert \\
   ~\eqwithref{\eqref{equ:four_matching_comparison}}{\leq}        & \alpha \left(\opt\left( \mathcal{S} \right) - \lvert M^{\max} \rvert\right) + 2 \lvert M^{\max} \rvert \\
   \eqnoref{=}                                                    & \alpha \opt\left(\mathcal{S} \right) + (2 - \alpha) \lvert M^{\max} \rvert \\
   ~\eqwithref{\eqref{equ:max_leq_s}}{\leq}                        & \alpha \opt\left(\mathcal{S} \right) + \frac{2 - \alpha}{2} \opt\left(\mathcal{S}\right) \\
   \eqnoref{=}                                                    & \frac{1 + \alpha}{2} \opt\left(\mathcal{S}\right).
   \end{align*}
   Thus, our algorithm has the desired approximation guarantee of~$\frac{1 + \alpha}{2}$.
\end{proof}

\subsection{Main algorithm for trees with little overlap}\label{subsec:four_variables_main_part}

It remains to deal with~$\Tmain$, which is the set of trees we build with depth~$2$.
\begin{lemma}\label{lem:opt_2_4_is_nice}
   There exists an optimum solution to instance~$\Tmain$ such the resulting circuit is of depth~2, i.e., that
   each node computing a Boolean function on four inputs has as predecessors nodes computing Boolean functions on two inputs\ifbool{long}{ (see \cref{fig:4_2_trafo})}.
\end{lemma}
\begin{proof}
   Since all trees that need to be built in the optimum solution cannot share more than two Boolean variables with any other tree,
   intermediate results computing a function on three inputs are never used more than once.
   If such a node~$v$ computes~$x_1 \circ x_2 \circ x_3$, and the tree actually wants to compute~$x_1 \circ x_2 \circ x_3 \circ x_4$, the last node~$u$ of the tree combines~$v$ with~$x_4$.
   Node~$v$ itself must have a predecessor~$w$ which computes an intermediate result, say without loss of generality~$x_1 \circ x_2$.
   Removing~$v$ and introducing a node~$w'$ computing~$x_3 \circ x_4$ instead and then computing~$u$ from~$w$ and~$w'$ does not increase the number of nodes but makes node~$u$ have depth~2.
   Hence, we can successively eliminate all trees of depth~3 and end up with the desired circuit.
\end{proof}

\ifbool{long}{
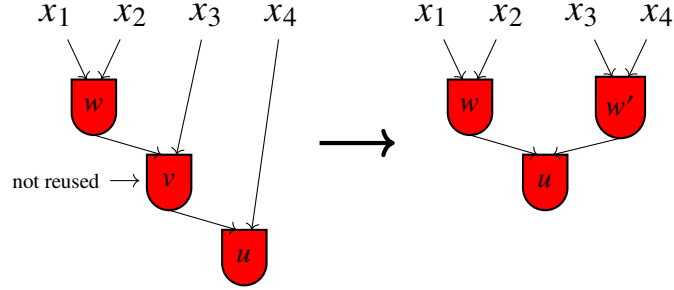
\begin{figure}[htbp]
   \begin{center}
      \begin{tikzpicture}
         \node[] at (1.5,1.0) (text) {\scriptsize not reused};

         \node[scale=1.3] (i1) at (1.5, 3.2){$x_1$};
         \node[scale=1.3] (i2) at (2.5, 3.2){$x_2$};
         \node[scale=1.3] (i3) at (3.5, 3.2){$x_3$};
         \node[scale=1.3] (i4) at (4.5, 3.2){$x_4$};

         \node[and-gate] at (2, 2)(node1){\rot{$w$}};
         \node[and-gate] at (3, 1)(node2){\rot{$v$}};
         \node[and-gate] at (4, 0)(node3){\rot{$u$}};

         \draw[->] (node1.output) -- (node2.input 2);
         \draw[->] (node2.output) -- (node3.input 2);
         \draw[->] (i1) -- (node1.input 2);
         \draw[->] (i2) -- (node1.input 1);
         \draw[->] (i3) -- (node2.input 1);
         \draw[->] (i4) -- (node3.input 1);

         \draw[->] (text) -- ($(node2.south) + (-0.1, 0)$);

         \draw[ultra thick, ->] (5, 1.5) -- (6, 1.5);

         \newcommand{\x}{5}
         \node[scale=1.3] (i1) at (1.5 + \x, 3.2){$x_1$};
         \node[scale=1.3] (i2) at (2.5 + \x, 3.2){$x_2$};
         \node[scale=1.3] (i3) at (3.5 + \x, 3.2){$x_3$};
         \node[scale=1.3] (i4) at (4.5 + \x, 3.2){$x_4$};

         \node[and-gate,] at (2 + \x, 2)(node4){\rot{$w$}};
         \node[and-gate,] at (4 + \x, 2)(node5){\rot{$w'$}};
         \node[and-gate,] at (3 + \x, 1)(node6){\rot{$u$}};
         \draw[->] (node4.output) -- (node6.input 2);
         \draw[->] (node5.output) -- (node6.input 1);
         \draw[->] (i1) -- (node4.input 2);
         \draw[->] (i2) -- (node4.input 1);
         \draw[->] (i3) -- (node5.input 2);
         \draw[->] (i4) -- (node5.input 1);

      \end{tikzpicture}
      \caption{We can apply this transformation to all trees in~$\Tfourtwo$ that are of depth~$3$ in the optimum solution. This does not change the optimality.}
      \label{fig:4_2_trafo}
   \end{center}
\end{figure}
}

We now use the hypergraph construction defined by~\eqref{equ:four_hypergraph_definition} and look for a vertex cover (compare~\crefrange{alg:step:four_graph_construction}{alg:step:four_solve_vc}).
It encodes our need to find the right depth 2 vertices as following.
The vertices of the hypergraph are pairs of indices and correspond to a pair of variables that we could combine by a gate.
For each vertex chosen in the vertex cover we will build a gate combining these two variables.

\begin{figure}[htbp]
   \begin{center}
      \begin{tikzpicture}
         \scriptsize
         \node[] (12) at (-0.2,4) {};
         \node[] (13) at (-1,2) {};
         \node[] (14) at (-0.2,0) {};
         \node[] (34) at (2.2,4) {};
         \node[] (24) at (3,2) {};
         \node[] (23) at (2.2,0) {};

         \begin{scope}
            \draw[] ($(12)+(0.3,1)$)
            to[out=0, in=90] ($(13) + (1.1,0)$)
            to[out=270, in=0] ($(14) + (0.3, -1)$)
            to[out=180, in=270] ($(13) + (-0.8, 0)$)
            to[out=90, in=180] ($(12) + (0.3, 1)$);

            \draw[color = red] ($(12)+(0.3,0.5)$)
            to[out=0, in=90] ($(13) + (0.8,0)$)
            to[out=270, in=90] ($(23) + (0.5, -0.1)$)
            to[out=270, in=0] ($(23) + (0, -0.5)$)
            to[out=180, in=315] ($(23) + (-0.5, -0.3)$)
            to[out=135, in=270] ($(13) + (-0.6, 0)$)
            to[out=90, in=180] ($(12) + (0.3, 0.5)$);

            \draw[color = blue] ($(14)+(0.3,-0.5)$)
            to[out=0, in=270] ($(13) + (0.6,0)$)
            to[out=90, in=270] ($(34) + (0.5, 0.1)$)
            to[out=90, in=0] ($(34) + (0, 0.5)$)
            to[out=180, in=45] ($(34) + (-0.5, 0.3)$)
            to[out=225, in=90] ($(13) + (-0.5, 0)$)
            to[out=270, in=180] ($(14) + (0.3, -0.5)$);

            \draw[color = orange] ($(34) + (0.7, 0)$)
            to[out=90, in=0] ($(34) + (0, 0.7)$)
            to[out=180, in=45] ($(34) + (-0.7, 0.4)$)
            to[out=225, in=90] ($(13) + (-0.7, 0)$)
            to[out=270, in=135] ($(23) + (-0.7, -0.4)$)
            to[out=315, in=180] ($(23) + (0, -0.7)$)
            to[out=0, in=270] ($(23) + (0.7, 0)$)
            to[out=90, in=270] ($(13) + (1.0,0)$)
            to[out=90, in=270] ($(34) + (0.7, 0)$);

            \draw[color = mid_green] ($(34)+(-0.3,1)$)
            to[out=180, in=90] ($(24) + (-1.1,0)$)
            to[out=270, in=180] ($(23) + (-0.3, -1)$)
            to[out=0, in=270] ($(24) + (0.8, 0)$)
            to[out=90, in=0] ($(34) + (-0.3, 1)$);

            \draw[color = cyan] ($(34)+(-0.3,0.5)$)
            to[out=180, in=90] ($(24) + (-0.8,0)$)
            to[out=270, in=90] ($(14) + (-0.5, -0.1)$)
            to[out=270, in=180] ($(14) + (0, -0.5)$)
            to[out=0, in=225] ($(14) + (0.5, -0.3)$)
            to[out=45, in=270] ($(24) + (0.6, 0)$)
            to[out=90, in=0] ($(34) + (-0.3, 0.5)$);

            \draw[color = yellow] ($(23)+(-0.3,-0.5)$)
            to[out=180, in=270] ($(24) + (-0.6,0)$)
            to[out=90, in=270] ($(12) + (-0.5, 0.1)$)
            to[out=90, in=180] ($(12) + (0, 0.5)$)
            to[out=0, in=135] ($(12) + (0.5, 0.3)$)
            to[out=315, in=90] ($(24) + (0.5, 0)$)
            to[out=270, in=0] ($(23) + (-0.3, -0.5)$);

            \draw[color = purple] ($(12) + (-0.7, 0)$)
            to[out=90, in=180] ($(12) + (0, 0.7)$)
            to[out=0, in=135] ($(12) + (0.7, 0.4)$)
            to[out=315, in=90] ($(24) + (0.7, 0)$)
            to[out=270, in=45] ($(14) + (0.7, -0.4)$)
            to[out=225, in=0] ($(14) + (0, -0.7)$)
            to[out=180, in=270] ($(14) + (-0.7, 0)$)
            to[out=90, in=270] ($(24) + (-1.0,0)$)
            to[out=90, in=270] ($(12) + (-0.7, 0)$);

         \end{scope}

         \node[] (12) at (-0.2,4) {$\{1, 2\}$};
         \node[] (13) at (-1,2) {$\{1,3\}$};
         \node[] (14) at (-0.2,0) {$\{1,4\}$};
         \node[] (34) at (2.2,4) {$\{3,4\}$};
         \node[] (24) at (3,2) {$\{2,4\}$};
         \node[] (23) at (2.2,0) {$\{2,3\}$};
      \end{tikzpicture}
      \caption{Hyperedges constructed for~$T = \{1, 2, 3, 4\} \in \Tfourtwo$ in \cref{alg:four_variables}.
      Each edge contains precisely one of the top vertices, one of the middle vertices and one of the bottom vertices.}
      \label{fig:four_hyperedges}
   \end{center}
\end{figure}
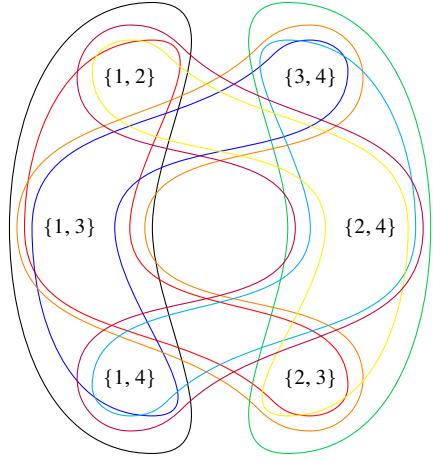

For each two variable tree we add a hyperedge that contains only the vertex consisting of this pair of variables.
This is equivalent to forcing to choose this vertex in the cover, which corresponds nicely to forcing to build a pair of variables.
Next, for each tree~$T$ with three variables, we add a hyperedge that contains all three two-element subsets of~$T$.
Like that, one of these sets is chosen by the vertex cover algorithm and we build one intermediate result for this tree.
Last but not least, we add hyperedges for trees with four variables.
This is a bit more complex.
We need to ensure that two disjoint subsets are chosen as vertices in the vertex cover.
Let~$T = \{1, 2, 3, 4\}$ without loss of generality.
Then for each combination of one of~$\{U_1^T, U_{\bar{1}}^T\}=\{\{1,2\}, \{3,4\}\}$, one of~$\{U_2^T, U_{\bar{2}}^T\}=\{\{1,3\}, \{2, 4\}\}$ and
one of~$\{U_3^T, U_{\bar{3}}^T\}=\{\{1,4\}, \{2,3\}\}$ we create one edge containing this combination (see \cref{fig:four_hyperedges}).
This ensures that if in each of these three sets one vertex is not chosen, then there is an uncovered edge.
So for one of these pairs both vertices need to be chosen.

\begin{lemma}\label{lem:vertex_cover_size}
   Let~$H$ be constructed as in \cref{alg:four_variables} \cref{alg:step:four_graph_construction}.
   The minimum size~$\opt_{\text{VC}}$ of a vertex cover in~$H$ is~$\opt(\Tmain) - \lvert \That \cup \Tfourtwo \rvert$.
\end{lemma}
\begin{proof}
   If you are given an optimum solution to~$\Tmain$, you can look at all pairs~$U$ of variables that are built in the optimum solution and take all these as a cover.
   For each~$T \in \Tfourtwo$ there exist sets~$U_{i}^T$ and~$U_{\bar{i}}^T$ such that both are built in the optimum solution.
   Now~$U_{i}^T$ and~$U_{\bar{i}}^T$ together cover all hyperedges associated with~$T$ since~$U_{i}^T \cup U_{\bar{i}}^T = T$.
   Similarly, for each~$T \in \That$ the intermediate result covers the hyperedge corresponding to~$T$.
   For each~$T \in \T_2$ the built vertex covers the one-element hyperedge.

   The number of such built pairs~$U$ is precisely~$\opt(\Tmain) - \lvert \That \cup \Tfourtwo \rvert$ because exactly~$\lvert \Tfourtwo \cup \That \rvert$ nodes of the optimum
   circuit are used to compute functions on three or four variables.
   Due to \cref{lem:opt_2_4_is_nice}, all other nodes compute functions on two variables.
   So
   \begin{align*}
      \opt_{\text{VC}} &\leq \opt(\Tmain) - \lvert \That \cup \Tfourtwo \rvert.
   \end{align*}

   Vice versa, look at the optimum vertex cover~$C$.
   The edges of~$H$ associated with a tree~$T$ are all covered by~$C$.
   For edges arising from a~$T \in \Tfourtwo$, this can only be true if for at least one~$i \in [3]$, we have~$U_{i}^T \in C$ and~$U_{\bar{\imath}}^T \in C$.
   Otherwise, there would be at most three~$U^T$ in~$C$ (for fixed~$T$) and then the edge containing precisely the other three~$U^T$ would not be covered.
   For edges arising from a~$T \in \That$ covering the edge directly yields the needed intermediate result.
   Regarding trees~$T \in \T_2$, this is clear, too.

   Hence, building nodes for each chosen~$U$ yields a set of nodes from which each~$T$ can be built using only (at most) one additional node.
   Thus, we also have
   \begin{align*}
      \opt_{\text{VC}} &\geq \opt(\Tmain) - \lvert \That \cup \Tfourtwo \rvert. \qedhere
   \end{align*}
\end{proof}

\begin{proposition}\label{pro:four_variables_depth_two}
   There exists a~$1.8$-approximation algorithm for~$\Tmain$.
\end{proposition}
\begin{proof}
   We use the reduction to \vc\ in hypergraphs as suggested in \cref{alg:four_variables}. \cref{alg:step:four_solve_vc}
   of \cref{alg:four_variables} gives us a vertex cover of size at most~$3 \opt_{\text{VC}}$.
   The reason for that is that there needs to be at least one vertex per matching edge and our solution uses up to three.
   We can be even more precise: The solution is of value at most~$3\opt_{\text{VC}} - 2 \lvert \T_2 \rvert$ because single vertex edges need to be covered in every solution.
   Computing a vertex cover in a graph or~$k$-hypergraph by using all vertices in a matching is a well-known strategy and with factor~$k$ the best known approximation for general \vc\ (see~\cite{dinur2003new}).
   After achieving this general guarantee, we want to improve our cover a bit further.
   Thus, in \cref{alg:step:four_variables_alternative_solution} we make sure that~$\lvert C \rvert \leq \lvert \T_2 \rvert + \lvert \That \rvert + 2 \lvert\Tfourtwo \rvert$, so in fact
   \begin{align}\label{equ:four_variables_size_vc}
      \lvert C \rvert \leq \min \left\{3 \opt_{\text{VC}} - 2 \lvert \T_2 \rvert, \lvert \T_2 \rvert + \lvert \That \rvert + 2 \lvert \Tfourtwo \rvert\right\}.
   \end{align}
   By \cref{lem:vertex_cover_size}, we know that the algorithm gives a solution of size at most~$C +\lvert \That \cup \Tfourtwo \rvert$, so
   \begin{align*}
                                                          & \frac{\alg(\Tmain)}{\opt(\Tmain)} \\
   ~\eqwithref{Lem.~\ref{lem:vertex_cover_size}}{=}       & \frac{\lvert \That \cup \Tfourtwo \rvert + \lvert C \rvert}{\lvert \That \cup \Tfourtwo \rvert + \opt_{\text{VC}}} \\
   ~\eqwithref{\eqref{equ:four_variables_size_vc}}{\leq}  & \frac{\lvert \That \cup \Tfourtwo \rvert + \min \left\{3 \opt_{\text{VC}} - 2 \lvert \T_2 \rvert, \lvert \T_2 \rvert + \lvert \That \rvert + 2 \lvert \Tfourtwo \rvert\right\}}
                                                                 {\lvert \That \cup \Tfourtwo \rvert + \opt_{\text{VC}}} \\
   \eqnoref{=}                                            & 1 + \frac{\min \left\{2 \opt_{\text{VC}} - 2 \lvert \T_2 \rvert, \lvert \T_2 \rvert + \lvert \That \rvert + 2 \lvert \Tfourtwo \rvert - \opt_{\text{VC}}\right\}}{\lvert \That \cup \Tfourtwo \rvert + \opt_{\text{VC}}} \\
   \eqnoref{\leq}                                         & 1 + \frac{\min \left\{2 \opt_{\text{VC}} - 2 \lvert \T_2 \rvert, 2 \lvert \Tmain \rvert - (\opt_{\text{VC}} - \T_2)\right\}}{\lvert \Tmain\rvert + \opt_{\text{VC}} - \lvert \T_2 \rvert} \\
   \eqnoref{=}                                            & 1 + \min \left\{2\frac{\opt_{\text{VC}} - \lvert \T_2 \rvert}{\lvert \Tmain\rvert + \opt_{\text{VC}} - \lvert \T_2 \rvert}
                                                            2 - 3\frac{\opt_{\text{VC}} - \lvert \T_2 \rvert }{\lvert \Tmain\rvert + \opt_{\text{VC}} - \lvert \T_2 \rvert}\right\} \\
   \eqnoref{\leq}                                         & 1 + \max_{x \in [0,1]} \min \{2x, 2 - 3x\} \\
   \eqnoref{=}                                            & \frac{9}{5}.
   \end{align*}
   This completes the proof.
\end{proof}

\begin{corollary}\label{cor:four_19}
   There exists a~$1.9$-approximation for~$\T \setminus (\Tsup \cup \Tintersect)$.
\end{corollary}
\begin{proof}
   Using the approximation guarantee of~$1.8$ given by~\cref{pro:four_variables_depth_two}, \cref{lem:four_matching} yields a guarantee of~$\frac{1 + 1.8}{2} = 1.9$.
\end{proof}

\begin{proof}[Proof of \cref{thm:main_result}~\ref{itm:k_4}]
   Putting the previous statements together, we can now prove \cref{thm:main_result}~\ref{itm:k_4}, i.e., that \cref{alg:four_variables} yields a~$1.9$-approximation.
   \ifbool{long}{
   \begin{align*}
                                                             & \alg(\T) \\
   \eqnoref{\leq}                                            & \alg\left(\T \setminus \left(\Tsup \cup \Tintersect\right)\right) + \frac 53 \lvert \Tintersect \rvert + \lvert \Tsup \rvert\\
   ~\eqwithref{Cor.~\ref{cor:four_19}}{\leq}                 & 1.9 \opt\left(\T \setminus (\Tsup \cup \Tintersect)\right)  + \frac 53 \lvert \Tintersect \rvert + \lvert \Tsup \rvert\\
   \eqnoref{\leq}                                            & 1.9 \opt\left(\T \setminus (\Tsup \cup \Tintersect)\right) + 1.9 \lvert \Tintersect \rvert + 1.9 \lvert \Tsup \rvert\\
   ~\eqwithref{\eqref{equ:remove_intersection}}{\leq}        & 1.9 \opt\left(\T \setminus \Tsup\right) + 1.9\lvert \Tsup \rvert \\
   ~\eqwithref{\eqref{equ:remove_tsup}}{=}                   & 1.9 \opt(\T)
   \end{align*}
   }

   \ifbool{short}{
      \begin{align*}
         & \alg(\T) \\
         \eqnoref{\leq}                                            & \alg\left(\T \setminus \left(\Tsup \cup \Tintersect\right)\right) + \frac 53 \lvert \Tintersect \rvert + \lvert \Tsup \rvert\\
         ~\eqwithref{Cor.~\ref{cor:four_19}}{\leq}                 & 1.9 \opt\left(\T \setminus (\Tsup \cup \Tintersect)\right)  + \frac 53 \lvert \Tintersect \rvert + \lvert \Tsup \rvert\\
         ~\eqwithref{\eqref{equ:remove_intersection}}{\leq}        & 1.9 \opt\left(\T \setminus \Tsup\right) + 1.9\lvert \Tsup \rvert \\
         ~\eqwithref{\eqref{equ:remove_tsup}}{=}                   & 1.9 \opt(\T)
      \end{align*}
   }

   The first inequality follows because having generated the tree for~$\T_2 \cup \That \cup \Tfourtwo$, we need precisely the following number of gates to add the other trees.
   \begin{itemize}
      \item For~$\Tsup$ we need exactly one gate per tree because we have build the three input subtree during the main part.
      \item For~$\Tintersect$ we need no more than~$\frac 53\lvert \Tintersect \rvert$ vertices in total.
      This is due to our definition of~$\Tintersect$, which is grouped in tuples each of which can be build with~$k+2$ vertices for~$k \geq 3$ trees.
   \end{itemize}
   This completes the proof.
\end{proof}

\section{General case with arbitrary many variables}\label{sec:arbitrary_variables}
Using some ideas we got from the small cases, we now want to give an approach to the general problem.
That is, we want to prove \cref{thm:main_result}~\ref{itm:general_k} from the introduction by analysing \cref{alg:bstso_size_up_to_k_2_3} and showing that it yields an efficient~$\frac 23 k$-approximation.

\renewcommand{\Tsup}{\T^{\supsetneq}}
\newcommand{\amount}{\alpha}

\begin{algorithm}
   \DontPrintSemicolon
   $\Tsup \coloneqq \{T \in \T : \exists T' \subsetneq T, T' \in \T, \lvert T' \rvert \geq \frac k3 \}$\;
   $\T = \T \setminus \Tsup$\;
   \For{$i \in \{2, \dots, \lvert \T \rvert\}$}
   {
      \While{$\exists S$ with~$\lvert S \rvert \geq \frac{i}{3(i-1)}k$ and~$S \subseteq T$ for at least~$i$ trees~$T \in \T$}
      {\label{alg:2_3size_k_reuse_start}
         Build~$S$ with~$\lvert S \rvert - 1$ vertices. \;
         Build each~$T \in \T$ with~$S \subseteq T$ using~$\lvert T \rvert - \lvert S \rvert$ vertices. \;
         $\T = (\T \setminus \{T : S \subseteq T\})$\;\label{alg:2_3size_k_reuse_end}
      }
   }
   Build each~$T \in \T$ on its own using~$\lvert T \rvert - 1$ vertices. \label{alg:2_3size_k_no_reuse} \;

   Build each~$T \in \Tsup$ reusing the vertices of~$T' \subsetneq T$. \label{alg:2_3size_k_supadding}\;

   \caption{\BSTSO\ algorithm for trees with up to~$k$ variables.}
   \label{alg:bstso_size_up_to_k_2_3}
\end{algorithm}

\cref{alg:bstso_size_up_to_k_2_3} is a generalization of what we have done in \cref{sec:three_variables}.
It works as follows.
In a preprocessing step, we store all those trees that can be built cheaply from other trees in~$\T$.
We call this set~$\Tsup$.
Basically, we run the algorithm without them and in \cref{alg:2_3size_k_supadding}, we build the stored trees from the main instance.

Now in the main part, for each number~$i$ we try to find subsets of at least~$i$ trees that overlap in a high enough number of variables that depends on~$i$.
If we find such a set~$S$, we directly build it and also build all trees containing~$S$.
Let the set of these trees be called~$\T'$.
We remove~$\T'$ from~$\T$ and continue.
Once we do not find such an~$S$ for any~$i$ anymore, we just build each remaining tree from scratch.
In the end, we build the trees in~$\Tsup$ from the already existing trees.

The analysis of \cref{alg:bstso_size_up_to_k_2_3} can be divided into three parts.
First, we briefly look at the guarantee of the preprocessing.
Second, we show that all the trees in~$\T'$ built in \cref{alg:2_3size_k_reuse_start} to \cref{alg:2_3size_k_reuse_end} need at most~$\frac 23 k \lvert \T' \rvert$ vertices in total,
which together with~$\opt \geq \lvert \T' \rvert$ gives the desired guarantee, too.
Last, we show that for the remaining trees of \cref{alg:2_3size_k_no_reuse}, the optimum solution is so bad that building them from scratch gives the desired guarantee.

\subsection{Analysis of the preprocessing}\label{subsec:arbitrary_preprocessing}
We make the following observations.
\begin{observation}\label{obs:preprocessing_1}
   Building each tree of~$\Tsup$ in the end takes at most~$\frac 23 k$ vertices because it is built from another tree containing at least~$\frac 13 k$ vertices.
\end{observation}
\begin{observation}\label{obs:preprocessing_2}
   In the optimum solution for the whole instance, there is at least on vertex per~$\T \in \Tsup$, namely the final vertex.
   Since there does not remain a tree~$T \in \T \setminus \Tsup$ which contains a tree in~$\Tsup$, all these~$\lvert \Tsup \rvert$ vertices are no longer needed and~$\opt (\T \setminus \Tsup) \leq \opt (\T) - \lvert \Tsup \rvert$.
\end{observation}
Putting these two observations together yields that we have a~$\frac 23 k$-approximation for the preprocessed trees.
Thus, it suffices to analyse the approximation guarantee of the main part of the algorithm with respect to the restricted instance~$\T \setminus \Tsup$.

\subsection{Analysis of the first part}\label{subsec:arbitrary_first_part}
We want to prove that we have a~$\frac 23k$-approximation for these trees.
\begin{lemma}\label{lem:general_first_part}
   The number of vertices used to build the~$i$ trees of a set~$\T'$ in~\crefrange{alg:2_3size_k_reuse_start}{alg:2_3size_k_reuse_end} is at most~$\frac 23 ik$.
\end{lemma}
\begin{proof}
   Let~$S$ be the intersection of all~$i$ in~$\T' \subseteq \T$.
   If we directly build this intersection~$S$ and then reuse it for all the trees~${T \in \T'}$, we save a lot of vertices compared to the worst case solution of having a single tree for each~${T \in \T'}$.
   By choosing a threshold of~$\frac{i}{3(i-1)}$ for the size of~$S$, the number of used vertices is bounded by
   \begin{align*}
      & \lvert S \rvert - 1 + \lvert \T' \rvert(k - \lvert S \rvert) \\
      = {}& i k - (i - 1) \lvert S \rvert \\
      \leq {}& i k - (i-1)\frac{i}{3(i - 1)}k \\
      = {}& \frac 23 ik. \qedhere
   \end{align*}
\end{proof}

Thus, given that~$\opt(\T \setminus \T') \leq \opt(\T) - \lvert \T' \rvert$, it suffices to prove the approximation guarantee for the remaining instance~$\T \setminus \T'$.

\subsection{Analysis of the second part}\label{subsec:arbitrary_second_part}
It is left to analyse the size of the optimum solution for the remaining instance.
We show that the optimum solution cannot be too cheap, implying that any solution already satisfies the wanted guarantee.

For a vertex~$v$, let~$a(v)$ be the number of trees that use vertex~$v$ in the optimum solution, where~$v$ is used by~$T$ if there is a~$v$-$v_T$ path for the output~$v_T$ computing~$T$.
We want to introduce a function saying how much of a vertex is owned by which tree.
First, for vertices~$v$ that correspond to variable sets~$S$ with less than~$\frac 13 k$ variables, we assign~$v$ completely to~$S$ if~$S \in \T$ or to no tree else.
For vertices~$v$ with larger variable set, we assign~$v$ by a fraction of~$\frac{1}{a(v)}$ to the trees~$T$ that use $v$.
Like that, each vertex is assigned at most once in total.

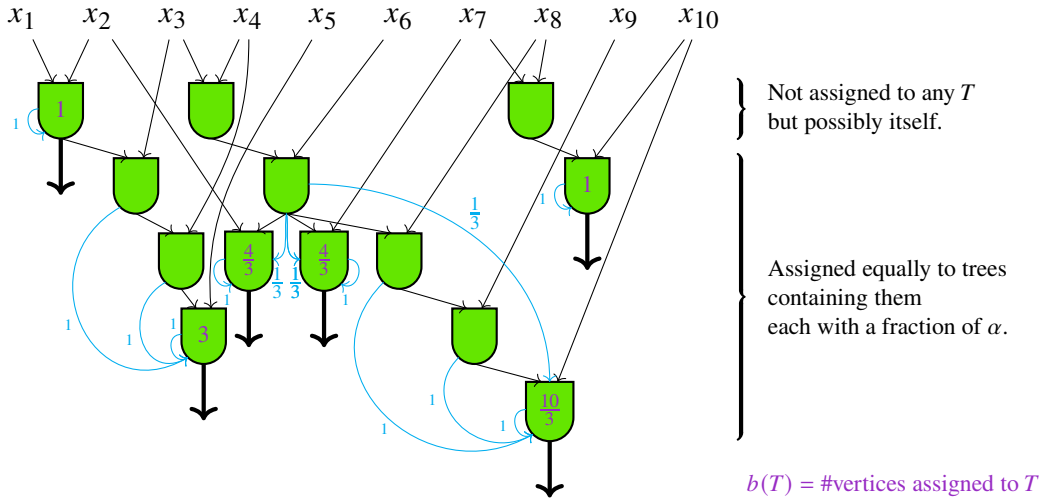
\begin{figure}[htbp]
   \begin{center}
      \begin{tikzpicture}
         \footnotesize
         \newcommand{\bvalue}[1]{\rot{\color{violet}$#1$}}
         \node[scale=1.3] (i1) at (1.5, 3.2){$x_1$};
         \node[scale=1.3] (i2) at (2.5, 3.2){$x_2$};
         \node[scale=1.3] (i3) at (3.5, 3.2){$x_3$};
         \node[scale=1.3] (i4) at (4.5, 3.2){$x_4$};
         \node[scale=1.3] (i5) at (5.5, 3.2){$x_5$};
         \node[scale=1.3] (i6) at (6.5, 3.2){$x_6$};
         \node[scale=1.3] (i7) at (7.5, 3.2){$x_7$};
         \node[scale=1.3] (i8) at (8.5, 3.2){$x_8$};
         \node[scale=1.3] (i9) at (9.5, 3.2){$x_9$};
         \node[scale=1.3] (i10) at (10.5, 3.2){$x_{10}$};

         \node[and-gate, fill=or_green] at (4,2) (and34) {};
         \node[and-gate, fill=or_green] at (5,1) (and346) {};
         \draw[->] (i3) -- (and34.input 2);
         \draw[->] (i4) -- (and34.input 1);
         \draw[->] (i6) -- (and346.input 1);
         \draw[->] (and34.output) -- (and346.input 2);

         \node[and-gate, fill=or_green] at (4.5, 0) (and2346) {\bvalue{\frac 43}};
         \draw[->] (i2) -- (and2346.input 2);
         \draw[->] (and346.output) -- (and2346.input 1);
         \draw[ultra thick, ->] (and2346) -- ($(and2346.output) + (0, -0.75)$);

         \node[and-gate, fill=or_green] at (5.5, 0) (and3467) {\bvalue{\frac 43}};
         \draw[->] (i7) -- (and3467.input 1);
         \draw[->] (and346.output) -- (and3467.input 2);
         \draw[ultra thick, ->] (and3467) -- ($(and3467.output) + (0, -0.75)$);

         \node[and-gate, fill=or_green] at (6.5, 0) (and3468) {};
         \draw[->] (i8) -- (and3468.input 1);
         \draw[->] (and346.output) -- (and3468.input 2);
         \node[and-gate, fill=or_green] at (7.5, -1) (and34689) {};
         \draw[->] (i9) -- (and34689.input 1);
         \draw[->] (and3468.output) -- (and34689.input 2);
         \node[and-gate, fill=or_green] at (8.5, -2) (and3468910) {\bvalue{\frac{10}{3}}};
         \draw[->] (i10) -- (and3468910.input 1);
         \draw[->] (and34689.output) -- (and3468910.input 2);
         \draw[ultra thick, ->] (and3468910) -- ($(and3468910.output) + (0, -0.75)$);

         \node[and-gate, fill=or_green] at (2,2) (and12) {\bvalue{1}};
         \node[and-gate, fill=or_green] at (3,1) (and123) {};
         \draw[->] (i1) -- (and12.input 2);
         \draw[->] (i2) -- (and12.input 1);
         \draw[->] (i3) -- (and123.input 1);
         \draw[->] (and12.output) -- (and123.input 2);
         \draw[ultra thick, ->] (and12) -- ($(and12.output) + (0, -0.75)$);

         \node[and-gate, fill=or_green] at (8.25,2) (and78) {};
         \node[and-gate, fill=or_green] at (9,1) (and7810) {\bvalue{1}};
         \draw[->] (i7) -- (and78.input 2);
         \draw[->] (i8) -- (and78.input 1);
         \draw[->] (i10) -- (and7810.input 1);
         \draw[->] (and78.output) -- (and7810.input 2);
         \draw[ultra thick, ->] (and7810) -- ($(and7810.output) + (0, -0.75)$);

         \node[and-gate, fill=or_green] at (3.6,0) (and1234) {};
         \node[and-gate, fill=or_green] at (3.9,-1) (and12345) {\bvalue{3}};
         \draw[->] (and123.output) -- (and1234.input 2);
         \draw[->] (i4) to[out=270, in=60] (and1234.input 1);
         \draw[->] (i5) to[out=240, in=95] (and12345.input 1);
         \draw[->] (and1234.output) -- (and12345.input 2);
         \draw[ultra thick, ->] (and12345) -- ($(and12345.output) + (0, -0.75)$);

         \draw[color=cyan, ->] (and12.south) to [out=180, in=210, looseness=2] node[midway, left] {\tiny 1} (and12.south east);
         \draw[color=cyan, ->] (and7810.south) to [out=180, in=210, looseness=2] node[midway, left] {\tiny 1} (and7810.south east);
         \draw[color=cyan, ->] (and123.south east) to [out=210, in=210, looseness=1.75] node[midway, left] {\tiny 1} (and12345.south east);
         \draw[color=cyan, ->] (and1234.south east) to [out=210, in=210, looseness=1.75] node[midway, left] {\tiny 1} (and12345.south east);
         \draw[color=cyan, ->] (and12345.south) to [out=180, in=210, looseness=2] node[pos=0.2, above] {\tiny 1} (and12345.south east);
         \draw[color=cyan, ->] (and346.east) to [out=260, in=0] node[pos=0.9, below] {\footnotesize $\frac 13$} (and2346.north);
         \draw[color=cyan, ->] (and2346.south) to [out=180, in=210, looseness=2] node[pos=0.9, below] {\tiny 1} (and2346.south east);
         \draw[color=cyan, ->] (and346.east) to [out=280, in=180] node[pos=0.9, below] {\footnotesize $\frac 13$} (and3467.south);
         \draw[color=cyan, ->] (and3467.north) to [out=0, in=-30, looseness=2] node[pos=0.9, below] {\tiny 1} (and3467.north east);
         \draw[color=cyan, ->] (and346.east) to [out=280, in=180] node[pos=0.9, below] {\footnotesize $\frac 13$} (and3467.south);
         \draw[color=cyan, ->] (and346.north) to [out=0, in=90] node[midway, above] {\footnotesize $\frac 13$} (and3468910.west);
         \draw[color=cyan, ->] (and3468.south east) to [out=225, in=210, looseness=1.5] node[midway, below] {\tiny 1} (and3468910.south east);
         \draw[color=cyan, ->] (and34689.south east) to [out=225, in=210, looseness=1.5] node[pos=0.3, left] {\tiny 1} (and3468910.south east);
         \draw[color=cyan, ->] (and3468910.south) to [out=180, in=210, looseness=2] node[midway, left] {\tiny 1} (and3468910.south east);

         \draw [ultra thick, decorate, decoration = {calligraphic brace}] (11,2.4) to node [label={[align=left]right:{Not assigned to any~$T$ \\  but possibly itself.}}, midway, right]
            {}  (11,1.6);
         \draw [ultra thick, decorate, decoration = {calligraphic brace}] (11,1.4) to node [label={[align=left]right:{Assigned equally to trees \\ containing them \\ each with a fraction of~$\alpha$.}}, midway, right]
            {}  (11,-2.4);

         \node[violet, anchor=west, align=left] (text) at (11, -3) {$b(T) = \#$vertices assigned to~$T$};
      \end{tikzpicture}
      \caption{Example instance for the analysis of the second part.
      In \textcolor{cyan}{blue} you can see which vertex is assigned to which tree by which fraction.
      These numbers are basically the~$\alpha(v)$ defined below.
      In \textcolor{violet}{purple} the total amount~$b(T)$ of vertices assigned to a tree~$T$ is shown (as a number within the output vertex).}
      \label{fig:general_analysis_2}
   \end{center}
\end{figure}
Let~$b(T)$ be the total amount of fractional vertices that is assigned to~$T$.
Then we have
\begin{align}\label{equ:opt_geq_bt}
\opt(\T) \geq \sum_{T \in \T} b(T).
\end{align}
Let in addition~$\amount(v) \coloneqq \frac{1}{a(v)}$ for vertices~$v$ representing sets of size at least~$\frac 13k$ and~$\amount(v) \coloneqq 0$ else.
So~$\amount(v)$ is the amount of~$v$ that is assigned to each~$T$ strictly containing~$v$.

\begin{lemma}\label{lem:b_per_tree}
   For each tree~$T$ left in~$\T$ in \cref{alg:2_3size_k_no_reuse}, we have
   \begin{align}\label{equ:b_per_tree}
      \frac{\lvert T \rvert}{b(T)} \leq \frac 2 3k.
   \end{align}
\end{lemma}
\begin{proof}
   Let~$v$ and~$w$ be the input vertices of the final gate of a tree~$T \in \T$.
   If either~$\amount(v) = 1$ or~$\amount(w) = 1$, then~$b(T) \geq 2$ and thus, the statement is clear.
   The same holds if~$\lvert T \rvert \leq \frac 23 k$ since~$b(T)\geq 1$ for all~$T$.

   Else, observe that~$b(T) \geq 1 + \amount(v) + \amount(w)$.
   In addition, we know that~$\lvert \T \rvert < \frac{1}{3(1 - \amount(v))} k + \frac{1}{3(1 -\amount(w))}k$ since otherwise,~$v$ or~$w$ would have been build in \cref{alg:2_3size_k_reuse_start} to \cref{alg:2_3size_k_reuse_end}.
   Putting this together
   \ifbool{long}{and using \cref{lem:technical_computation}, which we prove below, yields}

   \ifbool{short}{we get}

   \ifbool{long}{\begin{align}\label{equ:use_of_technical_lemma}
   & \frac{\lvert T \rvert }{b(T)}
   \eqnoref{\leq}  \frac{\frac{1}{3(1 - \amount(v))} + \frac{1}{3(1 - \amount(w))}}{1 + \amount(v) + \amount(w)}k
   ~\eqwithref{\eqref{equ:technical_computation}}{\leq}  \frac 23k.
   \end{align}
   }

   \ifbool{short}{\begin{align}\label{equ:use_of_technical_lemma}
   & \frac{\lvert T \rvert }{b(T)}
   \eqnoref{\leq}  \frac{\frac{1}{3(1 - \amount(v))} + \frac{1}{3(1 - \amount(w))}}{1 + \amount(v) + \amount(w)}k
   \eqnoref{\leq}  \frac 23k.
   \end{align}
   The second inequality is a technicality that is true for every~$\amount < \frac 12$.
   }
\end{proof}

\begin{lemma}\label{lem:general_second_part}
   The approximation guarantee of building everything from scratch in \cref{alg:2_3size_k_no_reuse} is~$\frac 23 k$.
\end{lemma}
\begin{proof}
   Summing~\eqref{equ:b_per_tree} over all trees left, we get
   \begin{align*}
   \frac{\sum_{T \in \T} \lvert T \rvert}{\opt}
   ~\eqwithref{\eqref{equ:opt_geq_bt}}{\leq} \frac{\sum_{T \in \T} \lvert T \rvert}{\sum_{T \in \T} b(T)}
   ~\eqwithref{\eqref{equ:frac_of_sums_leq_max_frac}}{\leq} \max_{T \in \T} \frac{\lvert T \rvert}{b(T)}
   ~\eqwithref{\eqref{equ:use_of_technical_lemma}}{\leq} \frac 23k.
   \end{align*}
   For the second step, we used the well-known formula
   \begin{align}\label{equ:frac_of_sums_leq_max_frac}
      \frac{\sum x_i}{\sum y_i} \leq \max \frac{x_i}{y_i}
   \end{align}
   for positive numbers~$x_i, y_i$.
   \ifbool{long}{A proof can be found below in \cref{lem:fraction_of_sums}. }

   \ifbool{short}{A proof can be found in the appendix. }

   This concludes the proof of the second part, i.e., that \cref{alg:2_3size_k_no_reuse} gives a~$\frac 23k$-approximation for the remaining instance.
\end{proof}
\begin{proof}[Proof of \cref{thm:main_result}~\ref{itm:general_k}]
   To prove the theorem, it remains to put everything together.
   \begin{align*}
      \alg(\T) \eqwithref{Ob.~\ref{obs:preprocessing_1}}{\leq} & \alg(\T \setminus \Tsup) + \frac 23 k \lvert \Tsup \rvert \\
      \eqwithref{Lem.~\ref{lem:general_first_part}}{\leq} & \alg(\{T \text{ built in \cref{alg:2_3size_k_no_reuse}}\}) + \frac 23 k \opt\left( \{T \text{ built in \crefrange{alg:2_3size_k_reuse_start}{alg:2_3size_k_reuse_end}}\} \right) +  \frac 23 k \lvert \Tsup \rvert \\
      \eqwithref{Lem.~\ref{lem:general_second_part}}{\leq} & \frac 23 k \opt(\T)
   \end{align*}
   Thus, we have shown that \cref{alg:bstso_size_up_to_k_2_3} gives an overall~$\frac 23 k$-approximation.
\end{proof}

\ifbool{long}{
It remains to prove the following technical ingredients.
\begin{lemma}\label{lem:technical_computation}
   Given two numbers~$\amount(v), \amount(w) \in \left[0,\frac 12\right]$, we have
   \begin{align}\label{equ:technical_computation}
      \frac{\frac{1}{3(1 - \amount(v))} + \frac{1}{3(1 - \amount(w))}}{1 + \amount(v) + \amount(w)}\leq \frac{2}{3}.
   \end{align}
\end{lemma}
\begin{proof}
   For every~$\amount \in \left[0, \frac 12\right]$ it holds that
   \begin{align}
      && 0 & \leq \amount - 2 \amount^2 \nonumber\\
      \iff && 1 & \leq 1 + \amount - 2 \amount^2 \nonumber\\
      \iff && 1 & \leq (1 + 2 \amount) (1 - \amount) \nonumber\\
      \iff && \frac{1}{1 - \amount} & \leq 1 + 2 \amount. \label{equ:basic_amount}
   \end{align}
   Adding~\eqref{equ:basic_amount} for~$\amount = \amount(v)$ and~$\amount = \amount(w)$ yields
   \begin{align*}
      && \frac{1}{1 - \amount(v)} + \frac{1}{1 - \amount(w)} & \leq 2 (1 + \amount(v) + \amount(w)) \\
      \iff && \frac{1}{3(1 - \amount(v))} + \frac{1}{3(1 - \amount(w))} & \leq \frac 23 (1 + \amount(v) + \amount(w))
   \end{align*}
   Dividing by~$1 + \amount(v) + \amount(w)$ yields the desired result~\eqref{equ:technical_computation}.
\end{proof}

Last but not least, we used the following well-known formula, which we want to prove for the sake of completeness.

\begin{lemma}\label{lem:fraction_of_sums}
   For non-negative real numbers~$x_1, \dots, x_n$ and positive numbers~$y_1, \dots, y_n$ we have
   \begin{align}\label{equ:fraction_of_sums}
      \frac{\displaystyle \sum_{i=1}^n x_i}{\displaystyle \sum_{i=1}^n y_i} \leq \max_{i=1}^n \frac{x_i}{y_i}.
   \end{align}
\end{lemma}
\begin{proof}
   Let~$M = \max_{i=1}^n \frac{x_i}{y_i}$.
   So~$x_i \leq M y_i$.
   Summing this over all~$i$ we get~$\sum_{i=1}^n x_i \leq M \sum_{i=1}^n{y_i}$
   which is equivalent to the desired inequality.
\end{proof}
}

\subsection*{Acknowledgments}
The author is grateful to Stephan Held and Martin Nägele for proofreading this paper as well as to Niklas Schlomberg for fruitful discussions.
\newpage
\begingroup
\setlength{\emergencystretch}{3em}
\printbibliography
\endgroup

\end{document}